\documentclass[twocolumn,showpacs,preprintnumbers,amsmath,amssymb,superscriptaddress,nofootinbib,english]{revtex4-1}
\usepackage{times,amsmath,amsfonts,amssymb,epstopdf}
\usepackage{graphicx}
\usepackage{dcolumn}
\usepackage{bm}
\usepackage{epsfig}
\usepackage{graphicx}
\usepackage{hyperref}
\usepackage[usenames]{color}
\usepackage{url}
\usepackage[normalem]{ulem}
\usepackage[T1]{fontenc}

\def\be{\begin{equation}}
\def\ee{\end{equation}}
\def\ben{\begin{eqnarray}}
\def\een{\end{eqnarray}}
\def\ba{\begin{array}}
\def\ea{\end{array}}

\newcommand{\bq}{\begin{eqnarray}}
\newcommand{\eq}{\end{eqnarray}}
\newcommand{\bes}{\begin{subequations}}
\newcommand{\ees}{\end{subequations}}

\begin{document}
\newcommand{\half}{{\textstyle\frac{1}{2}}}
\allowdisplaybreaks[3]
\def\triangledown{\nabla}
\def\grad3{\hat{\nabla}}
\def\a{\alpha}
\def\b{\beta}
\def\g{\gamma}\def\G{\Gamma}
\def\d{\delta}\def\D{\Delta}
\def\ep{\epsilon}
\def\et{\eta}
\def\z{\zeta}
\def\t{\theta}\def\T{\Theta}
\def\l{\lambda}\def\L{\Lambda}
\def\m{\mu}
\def\f{\phi}\def\F{\Phi}
\def\n{\nu}
\def\p{\psi}\def\P{\Psi}
\def\r{\rho}
\def\s{\sigma}\def\S{\Sigma}
\def\ta{\tau}
\def\x{\chi}
\def\o{\omega}\def\O{\Omega}
\def\k{\kappa}
\def\pa {\partial}
\def\ov{\over}
\def\br{\\}
\def\ud{\underline}

\newcommand\lsim{\mathrel{\rlap{\lower4pt\hbox{\hskip1pt$\sim$}}
    \raise1pt\hbox{$<$}}}
\newcommand\gsim{\mathrel{\rlap{\lower4pt\hbox{\hskip1pt$\sim$}}
    \raise1pt\hbox{$>$}}}
\newcommand\esim{\mathrel{\rlap{\raise2pt\hbox{\hskip0pt$\sim$}}
    \lower1pt\hbox{$-$}}}
\newcommand{\dpar}[2]{\frac{\partial #1}{\partial #2}}
\newcommand{\sdp}[2]{\frac{\partial ^2 #1}{\partial #2 ^2}}
\newcommand{\dtot}[2]{\frac{d #1}{d #2}}
\newcommand{\sdt}[2]{\frac{d ^2 #1}{d #2 ^2}}    

\title{The parameter space in Galileon gravity models}

\author{Alexandre Barreira}
\email[Electronic address: ]{a.m.r.barreira@durham.ac.uk}
\affiliation{Institute for Computational Cosmology, Department of Physics, Durham University, Durham DH1 3LE, U.K.}
\affiliation{Institute for Particle Physics Phenomenology, Department of Physics, Durham University, Durham DH1 3LE, U.K.}

\author{Baojiu Li}
\affiliation{Institute for Computational Cosmology, Department of Physics, Durham University, Durham DH1 3LE, U.K.}

\author{Ariel Sanchez}
\affiliation{Max-Planck-Institut f\"{u}r extraterrestrische Physik, Postfach 1312, Giessenbachstr., 85748 Garching, Germany}

\author{Carlton M. Baugh}
\affiliation{Institute for Computational Cosmology, Department of Physics, Durham University, Durham DH1 3LE, U.K.}

\author{Silvia Pascoli}
\affiliation{Institute for Particle Physics Phenomenology, Department of Physics, Durham University, Durham DH1 3LE, U.K.}

\begin{abstract}

We present the first constraints on the full parameter space of the Galileon modified gravity model, considering both the cosmological parameters and the coefficients which specify the additional terms in the Lagrangian due to the Galileon field, which we call the Galileon parameters.  We use the latest cosmic microwave background measurements, along with distance measurements from supernovae and baryonic acoustic oscillations, performing a Monte Carlo Markov Chain exploration of the 9-dimensional parameter space. The integrated Sachs-Wolfe signal can be very different in Galileon models compared to standard gravity, making it essential to use the full CMB data rather than the CMB distance priors. We demonstrate that meaningful constraints are only possible in the Galileon parameter space after taking advantage of a scaling degeneracy. We find that the Galileon model can fit the WMAP 9-year results {\it better} than the standard $\Lambda$-Cold Dark Matter model, but gives a slightly worse fit overall once lower redshift distance measurements are included.  The best-fitting cosmological parameters (e.g. matter density, scalar spectral index, fluctuation amplitude) can differ by more than $2\sigma$ in the Galileon model compared with $\Lambda$CDM.  We highlight other potential constraints of the Galileon model using galaxy clustering and weak lensing measurements.

\end{abstract} 
\maketitle

\section{Introduction}

One of the major unsolved mysteries in cosmology is the nature of the `dark energy' that is causing the observed present-day accelerated expansion of the Universe \cite{Guy:2010bc, Percival:2009xn, Beutler:2011hx, Reid:2012sw, Anderson:2012sa, Sanchez:2012sg, Hinshaw:2012fq, Riess:2009pu, Riess:2011yx, Freedman:2012ny}. In the context of General Relativity (GR), the vacuum energy, also known as the cosmological constant, $\Lambda$, is the simplest explanation for dark energy. In the current standard cosmology, known as the $\Lambda$ - Cold Dark Matter ($\Lambda$CDM) model, the cosmological constant accounts for approximately 70$\%$ of the total energy content of the Universe today, with non-relativistic dark and baryonic matter, neutrinos and photons completing the cosmic inventory. $\Lambda$CDM performs very well against observations. However, this model has serious theoretical problems, the biggest one being the huge discrepancy of many orders of magnitude between the standard quantum field prediction for $\Lambda$ and the value inferred from observations. It has become widely accepted that one needs to look beyond $\Lambda$. A possible and attractive alternative is to modify GR on large scales in such a way to allow cosmic acceleration without the need to invoke a cosmological constant or any other negative pressure fluid \cite{2006IJMPD..15.1753C, Li:2011sd}. Interest in modified gravity models has been growing over the past decade \cite{Clifton:2011jh}, with significant progress being made in both analytical modelling \cite{Brax:2011aw, Brax:2012gr, Baker:2012zs, Baker:2011jy, Battye:2012eu, Pearson:2012kb, Bloomfield:2012ff, Gubitosi:2012hu, Sawicki:2012re} and numerical simulations \cite{Li:2011vk, Jennings:2012pt, Li:2012by}.

Long before the discovery of the accelerated expansion, in 1974, Horndeski derived the most general action for a single scalar field that yields equations of motion (EOM) which contain up to second order derivatives of the fields \cite{Horndeski:1974wa}, avoiding, therefore, the presence of Ostrogradsky ghost degrees of freedom \cite{Woodard:2006nt}. However, it was only recently that such an action started to gain relevance in cosmology \cite{Deffayet:2011gz, Kobayashi:2011nu}. A large number of successful modified gravity models can be considered as special cases of the general Horndeski theory, which satisfy certain desired properties or symmetries. The most famous recent examples include the Galileon model \cite{PhysRevD.79.064036, PhysRevD.79.084003, Deffayet:2009mn}, kinetic gravity braiding \cite{Deffayet:2010qz, Pujolas:2011he, Kimura:2011td}, Fab-Four \cite{Charmousis:2011bf, Charmousis:2011ea, Bruneton:2012zk, Copeland:2012qf}, Fab-Five \cite{Appleby:2012rx}, k-mouflage \cite{Babichev:2009ee} and others \cite{Kobayashi:2009wr, Kobayashi:2010wa, Leon:2012mt, Silva:2009km}. 

In this paper we focus on the Galileon model, which was originally derived by generalizing the four-dimensional effective boundary action of the braneworld DGP model \cite{2000PhLB..485..208D, 2003JHEP...09..029L, 2004JHEP...06..059N, deRham:2012az}. In both these models, the modifications to gravity are determined by an extra scalar degree of freedom $\varphi$, known as the Galileon, whose Lagrangian is invariant under the Galilean shift symmetry $\partial_\mu\varphi \rightarrow \partial_\mu\varphi + b_\mu$, where $b_\mu$ is a constant vector. The branch of the DGP model that leads to cosmic acceleration suffers from ghosts \cite{1126-6708-2003-09-029, 1126-6708-2004-06-059, PhysRevD.73.044016}; the Galileon is free from such problems. In a four-dimensional Minkowski space there are only five Galilean invariant Lagrangians that are second order in the equations of motion, despite containing highly nonlinear derivative couplings of the scalar field \cite{PhysRevD.79.064036}. In order to investigate the cosmological implications of the Galileon model, its Lagrangian needs to be generalized to curved spacetimes and in \cite{PhysRevD.79.084003, Deffayet:2009mn} the authors have shown that explicit couplings between the Galileon field derivatives and the curvature tensors are needed to retain the equations of motion up to second order, although they explicitly break the Galilean symmetry.

The presence of the nonlinear derivative interactions and the couplings to the curvature tensors in the Galileon model change the way in which matter determines the geometry of spacetime relative to the GR prediction. However, near massive objects like the Sun, the laws of gravity are constrained to be very close to GR and, therefore, any viable modified gravity model needs to revert to the `standard' solution on these scales \cite{1990PhRvL..64..123D, 1999PhRvL..83.3585B, Will:2005va}. Remarkably, the nonlinear derivative terms allow, at the same time, for the suppression (or screening) of the deviations from GR by effectively decoupling the scalar field from gravity in high curvature regions by a mechanism known as Vainshtein screening \cite{Vainshtein1972393}. In brief, near matter sources, the nonlinear terms of the equations of motion become important, strongly suppressing the spatial gradient of the scalar field, which is the extra `fifth force', on scales below the so-called `Vainshtein radius'. On the other hand, beyond the Vainshtein radius the nonlinear terms are subdominant, such that both the Galileon field and the Newtonian potential satisfy a linear Poisson equation, and the total gravitational force gets a non-negligible contribution from the `fifth' and the `normal' GR forces. The Vainshtein mechanism is crucial to the success of the Galileon model and it is in many respects similar to the implementation of other screening mechanisms such as the chameleon \cite{PhysRevD.69.044026, PhysRevD.70.123518}, dilaton \cite{Brax:2010gi, Brax:2011aw, Brax:2012gr}, symmetron \cite{Hinterbichler:2011ca, Hinterbichler:2010es, Davis:2011pj, Brax:2011pk} and disformal screening \cite{Koivisto:2012za}. The main difference is that in the case of the Vainshtein mechanism, the range of the screening depends only on the properties of the gravitational source as opposed to the chameleon mechanism, in which case the screening depends also on the cosmological environment. 

The way the background cosmological evolution can be affected in Galileon models has already been studied and constrained using observational probes of the geometry of the Universe such as type Ia Supernovae (SNIa), the scale of the Baryonic Acoustic Oscillations (BAO) feature in the galaxy distribution and the position of the peaks of the Cosmic Microwave Background (CMB) angular power spectrum of temperature fluctuations \cite{Gannouji:2010au, PhysRevD.80.024037, DeFelice:2010pv, Nesseris:2010pc, Appleby:2011aa, PhysRevD.82.103015, Neveu:2013mfa}. Furthermore, by studying the evolution of density fluctuations in linear perturbation theory, it has also been shown how Galileon models can be constrained theoretically, by the imposition of conditions to avoid the presence of ghosts and other pathologies \cite{DeFelice:2010pv, Appleby:2011aa}, and observationally, by investigating the predictions for the growth rate of Large Scale Structure (LSS) \cite{Appleby:2012ba, Okada:2012mn, PhysRevD.82.103015, Neveu:2013mfa} and for the full CMB, linear matter and weak lensing potential power spectra \cite{Barreira:2012kk} (see also \cite{Bartolo:2013ws} for a study of the bispectrum of the matter fluctuations). These studies revealed a number of distinctive features of the Galileon model, which could, in principle, be used to place strong constraints on the model and distinguish it from the $\Lambda$CDM paradigm and from other modified gravity theories.

In \cite{Barreira:2012kk} we took the first step towards the thorough exploration of the Galileon parameter space that we present in this paper. In particular, by using a modified version of the publicly available {\tt CAMB} code \cite{camb_notes}, we showed that the Galileon model predictions depend sensitively on the values chosen for the model parameters. We demonstrated that the Integrated Sachs-Wolfe (ISW) effect can play a key role in constraining the Galileon parameter space, thanks to the distinctive signature the model has on the largest angular scales of the CMB power spectrum. Our previous work has therefore motivated the analysis presented in this paper, where we explore the full cosmological parameter space and not only its Galilean subspace, using Monte Carlo Markov Chain methods (MCMC) with the aid of the publicly available {\tt CosmoMC} code \cite{Lewis:2002ah}. To allow the rest of the cosmological parameters to vary is crucial to fully understand the degeneracies that might exist between them, and it is, ultimately, the only way to truly understand how well a given model is able to reproduce the observational data.

The present paper is structured as follows. In Section \ref{The model} we introduce the covariant Galileon model, present the fully covariant and gauge invariant perturbed field equations we derived in \cite{Barreira:2012kk} and that we use to obtain the results. We also recap the main physical features of the model. In Section \ref{Methodology} we specify and discuss the full cosmological parameter space which we constrain, the theoretical priors that determine the viability of some regions of the parameter space and the datasets that we use to derive the constraints. In particular, we point out that if all of the parameters that enter the model Lagrangian are allowed to vary, then there is an infinite degeneracy region along the Galilean subset of the full parameter space. We present and analyse the results in Section \ref{Results} where we look at how different combinations of data constrain the Galileon model, highlighting the tensions between the best-fitting models derived from different datasets. We present our conclusions in Section \ref{Conclusion}.

Throughout this paper we assume the metric convention $(+,-,-,-)$ and work in units in which the speed of light $c = 1$. Greek indices run over $0,1,2,3$ and we use $8\pi G=\kappa=M^{-2}_{\rm Pl}$ interchangeably, where $G$ is Newton's constant and $M_{\rm Pl}$ is the reduced Planck mass.

\section{The Galileon model}

\label{The model}

In this section we briefly review the covariant Galileon model and the field equations used to obtain the physical predictions of the model. For a detailed derivation we refer the reader to \cite{Barreira:2012kk} (see also \cite{PhysRevD.82.124054}).

The full action of the covariant Galileon model, which has no direct coupling between matter and the Galileon field (see however \cite{Appleby:2011aa,  Sushkov:2009hk, Gubitosi:2011sg, deRham:2011by, VanAcoleyen:2011mj, Zumalacarregui:2012us, Amendola:1993uh, Barreira:2012kk}), is given by

\bq\label{Galileon action}
&& S = \int d^4x\sqrt{-g} \left[ \frac{R}{16\pi G} - \frac{1}{2}\sum_{i=1}^5c_i\mathcal{L}_i - \mathcal{L}_m\right],
\eq
where $g$ is the determinant of the metric $g_{\mu\nu}$, $R$ is the Ricci scalar and the model parameters $c_{1-5}$ are dimensionless constants.
The five terms in the Lagrangian density, fixed by the Galilean invariance in flat spacetime, $\partial_\mu\varphi \rightarrow \partial_\mu\varphi + b_\mu$, are given by
\bq\label{L's}
\mathcal{L}_1 &=& M^3\varphi, \nonumber \\
\mathcal{L}_2 &=& \nabla_\mu\varphi\nabla^\mu\varphi,  \nonumber \\
\mathcal{L}_3 &=& \frac{2}{M^3}\Box\varphi\nabla_\mu\varphi\nabla^\mu\varphi, \nonumber \\
\mathcal{L}_4 &=& \frac{1}{M^6}\nabla_\mu\varphi\nabla^\mu\varphi\Big[ 2(\Box\varphi)^2 - 2(\nabla_\mu\nabla_\nu\varphi)(\nabla^\mu\nabla^\nu\varphi) \nonumber \\
&& -R\nabla_\mu\varphi\nabla^\mu\varphi/2\Big], \nonumber \\
\mathcal{L}_5 &=&  \frac{1}{M^9}\nabla_\mu\varphi\nabla^\mu\varphi\Big[ (\Box\varphi)^3 - 3(\Box\varphi)(\nabla_\mu\nabla_\nu\varphi)(\nabla^\mu\nabla^\nu\varphi) \nonumber \\
&& + 2(\nabla_\mu\nabla^\nu\varphi)(\nabla_\nu\nabla^\rho\varphi)(\nabla_\rho\nabla^\mu\varphi) \nonumber \\
&& -6 (\nabla_\mu\varphi)(\nabla^\mu\nabla^\nu\varphi)(\nabla^\rho\varphi)G_{\nu\rho}\Big],
\eq
in which $\varphi$ is the Galileon scalar field and $M^3\equiv M_{\rm Pl}H_0^2$ with $H_0$ being the present-day Hubble expansion rate. The derivative couplings to the Ricci scalar $R$ and the Einstein tensor $G_{\mu\nu}$ in $\mathcal{L}_4$ and $\mathcal{L}_5$ are necessary to prevent the EOMs from having higher than second-order derivatives of the metric and the Galileon field in curved spacetimes, such as the one described by the Friedman-Robertson-Walker (FRW) metric \cite{PhysRevD.79.084003}. Such terms, however, break the Galilean shift symmetry. 

As is customary in Galileon studies, one is generally interested in cases where cosmic acceleration is due only to the field kinetic terms and therefore we will set the potential term $c_1$ to zero. Note also that an important difference between the Galileon model and other subsets of Horndeski theory is that there are no free fuctions in the action of Eq.~(\ref{Galileon action}). In particular, this prevents the Galileon model from having a $\Lambda$CDM limit which in the end plays an important role in distinguishing between these two models (see however \cite{DeFelice:2010nf, DeFelice:2011aa}).

The modified Einstein equations and the Galileon EOM are obtained by varying the action of Eq.~(\ref{Galileon action}), with respect to $g_{\mu\nu}$ and $\varphi$, respectively. Given the length of these equations we do not show them in this paper. Interested readers can find them in Eqs.~(A1--A7) of Ref.~\cite{Barreira:2012kk}.

\subsection{Friedmann equations}

In the Galileon model, the structure of the two Friedmann equations is not changed with respect to standard GR. In this paper we will always consider the case of a spatially flat Universe for which the first and second Friedmann equations are given by:

\bq\label{background1}
\frac{1}{3}\theta^2 = \kappa {\bar{\rho}}, \\
\label{background2}
\dot{\theta} + \frac{1}{3}\theta^2 + \frac{\kappa}{2}({\bar{\rho}} + 3{\bar{p}}) &=& 0,
\eq
where $\theta = 3\dot{a}/a = 3H$, where $a$ is the cosmological scale factor, $H$ the Hubble expansion rate (throughout the paper we shall use $\theta$ and $H$ interchangeably); the overbar denotes background averaged quantities, the overdot denotes a derivative with respect to cosmic time and we are preserving the notation used in \cite{Barreira:2012kk}. 

In Eqs.~(\ref{background1}) and (\ref{background2}), $\bar{p} = \bar{p}_r + \bar{p}_m + \bar{p}_\varphi$ and $\bar{\rho} = \bar{\rho}_r + \bar{\rho}_m + \bar{\rho}_\varphi$ are, respectively, the total pressure and energy density in the Universe. The subscripts $_r$, $_m$ and $_\varphi$ denote the contribution from the relativistic degrees of freedom (photons and massless neutrinos), non-relativistic matter (dark and baryonic) and the Galileon field, respectively. In the case of the uncoupled Galileon model, the cosmic evolution of $\bar{\rho}_r$ and $\bar{\rho}_m$ is fixed by their present-day values $\bar{\rho}_{r0}$ and $\bar{\rho}_{m0}$, as

\bq
\label{radiation-matter-density}\bar{\rho}_r = \bar{\rho}_{r0} a^{-4} = 3H_0^2\kappa^{-1}\Omega_{r}^0 a^{-4} , \nonumber \\
\bar{\rho}_m = \bar{\rho}_{m0} a^{-3} = 3H_0^2 \kappa^{-1}\Omega_{m}^0 a^{-3} ,
\eq
 where $\Omega_{r}^0$ and $\Omega_{m}^0$ are the present-day values of the fractional energy density of the radiation and matter components, respectively. The background pressure of the radiation and matter is given, respectively by,

\bq\label{radiation-matter-pressure}
\bar{p}_r &=& \bar{\rho}_r / 3 \nonumber, \\
\bar{p}_m  &\approx& 0.
\eq

\subsection{Galileon field equations}

To describe the Galileon field up to linear order in perturbations, we shall make $3+1$ spacetime decompositions of the tensor quantities \cite{PhysRevD.40.1804, Challinor:1998xk}. This is achieved by using the projection tensor $h_{\mu\nu}$, which is defined as $h_{\mu\nu} = g_{\mu\nu} - u_\mu u_\nu$ and can be used to obtain projected covariant tensors which live in 3-dimensional hyperspaces perpendicular to an observer's 4-velocity $u^\mu$. For example, the covariant spatial derivative $\hat{\nabla}$ of a tensor field $T^{\beta...\gamma}_{\sigma...\lambda}$ is defined as
\bq
\hat{\nabla}^\alpha T^{\beta\cdot\cdot\cdot\gamma}_{\sigma\cdot\cdot\cdot\lambda} \equiv h^{\alpha}_{\mu}h^{\beta}_{\nu}\cdot\cdot\cdot h^{\gamma}_{\kappa}h^{\rho}_{\sigma}\cdot\cdot\cdot h^{\eta}_{\lambda}\nabla^\mu T^{\nu\cdot\cdot\cdot\kappa}_{\rho\cdot\cdot\cdot\eta}.
\eq

The energy-momentum tensor and the covariant derivative of the observer's 4-velocity field $u^\mu$ can be decomposed, respectively, as 

\bq
\label{Tuv} T_{\mu\nu} &=& \pi_{\mu\nu} + 2q_{(\mu}u_{\nu)} + \rho u_\mu u_\nu - ph_{\mu\nu},\\
\nabla_\mu u_\nu &=& \sigma_{\mu\nu} + \varpi_{\mu\nu} + \frac{1}{3}\theta h_{\mu\nu} + u_\mu A_\nu,
\eq
where $\pi_{\mu\nu}$ is the projected symmetric and trace-free (PSTF) anisotropic stress, $q_\mu$ is the heat flux vector, $p$ is the isotropic pressure, $\rho$ is the energy density, $\sigma_{\mu\nu}$ the PSTF shear tensor, $\varpi_{\mu\nu} = \hat{\nabla}_{[\mu}u_{\nu]}$ the vorticity, $\theta = \nabla^\alpha u_\alpha = 3H$ as introduced above and $A_\mu = \dot{u}_\mu$ is the observer's 4-acceleration. The time derivative can be expressed in terms of the covariant derivatives as $\dot{\phi} = u^\alpha\nabla_\alpha\phi$. Square brackets mean antisymmetrization and parentheses symmetrization. The normalization is $u^\alpha u_\alpha = 1$, in accordance with our choice for the metric signature. 

The physical quantities in the total energy-momentum tensor $T_{\mu\nu}$, Eq.~(\ref{Tuv}), have contributions from each of the matter species in the universe: 

\bq
\label{physical quantities 1} \rho &=& \rho^r + \rho^m + \rho^\varphi,\\
\label{physical quantities 2} p &=& p^r + p^m + p^\varphi, \\ 
\label{physical quantities 3} q_\mu &=& q_\mu^r + q_\mu^m + q_\mu^\varphi, \\
\label{physical quantities 4} \pi_{\mu\nu} &=& \pi_{\mu\nu}^r + \pi_{\mu\nu}^m + \pi_{\mu\nu}^\varphi .
\eq

In \cite{Barreira:2012kk} we presented a detailed explanation of how to derive the Galileon field contribution to the quantities of Eqs.~(\ref{physical quantities 1} - \ref{physical quantities 4}). However, for brevity,  in this paper we limit ourselves to quoting the final result, which, up to first order in perturbed quantities, is given by

\begin{widetext}
\bq\label{perturbed1}
\rho^\varphi &\doteq& c_2\left[ \frac{1}{2}\dot{\varphi}^2\right] + \frac{c_3}{M^3} \left[ 2\dot{\varphi}^3\theta + 2\dot{\varphi}^2\hat{\Box}\varphi\right]  +\frac{c_4}{M^6}\left[ \frac{5}{2}\dot{\varphi}^4\theta^2 + 4\dot{\varphi}^3\theta\hat{\Box}\varphi + \frac{3}{4}\dot{\varphi}^4\hat{R}\right]     \nonumber \\
&& + \frac{c_5}{M^9}\left[\frac{7}{9}\dot{\varphi}^5\theta^3 + \frac{5}{3}\dot{\varphi}^4\theta^2\hat{\Box}\varphi +\frac{1}{2}\dot{\varphi}^5\theta\hat{R} \right], \\
\label{perturbed2}
p^\varphi &\doteq& c_2\left[ \frac{1}{2}\dot{\varphi}^2\right] + \frac{c_3}{M^3} \left[ -2\ddot{\varphi}\dot{\varphi}^2\right] \nonumber \\
&&  + \frac{c_4}{M^6}\left[ -4\ddot{\varphi}\dot{\varphi}^3\theta - \dot{\varphi}^4\dot{\theta} - \frac{1}{2}\dot{\varphi}^4\theta^2 - 4\ddot{\varphi}\dot{\varphi}^2\hat{\Box}\varphi - \frac{4}{9}\dot{\varphi}^3\theta\hat{\Box}\varphi + \dot{\varphi}^4\hat{\triangledown}\cdot A + \frac{1}{12}\dot{\varphi}^4\hat{R}\right] \nonumber \\
&& + \frac{c_5}{M^9}\left[-\frac{5}{3}\ddot{\varphi}\dot{\varphi}^4\theta^2 - \frac{2}{3}\dot{\varphi}^5\dot{\theta}\theta - \frac{2}{9}\dot{\varphi}^5\theta^3 - \frac{2}{9}\dot{\varphi}^4\theta^2\hat{\Box}\varphi - \frac{8}{3}\ddot{\varphi}\dot{\varphi}^3\theta\hat{\Box}\varphi - \frac{1}{2}\ddot{\varphi}\dot{\varphi}^4\hat{R} - \frac{2}{3}\dot{\varphi}^4\dot{\theta}\hat{\Box}\varphi + \frac{2}{3}\dot{\varphi}^5\theta\hat{\triangledown} \cdot A\right],\\
\label{perturbed3}
q_\mu^\varphi &\doteq& c_2\left[ \dot{\varphi}\hat{\triangledown}_\mu\varphi \right] + \frac{c_3}{M^3} \left[ 2\dot{\varphi}^2\theta\hat{\triangledown}_\mu\varphi - 2\dot{\varphi}^2\hat{\triangledown}_\mu\dot{\varphi}\right] \nonumber \\
&&  + \frac{c_4}{M^6}\left[ -4\dot{\varphi}^3\theta\hat{\triangledown}_\mu\dot{\varphi} + 2\dot{\varphi}^3\theta^2\hat{\triangledown}_\mu\varphi - \dot{\varphi}^4\hat{\triangledown}_\mu\theta + \frac{3}{2}\dot{\varphi}^4\hat{\triangledown}^\alpha\sigma_{\mu\alpha} + \frac{3}{2}\dot{\varphi}^4\hat{\triangledown}^\alpha{\varpi}_{\mu\alpha} \right] \nonumber \\
&& + \frac{c_5}{M^9}\left[ - \frac{5}{3}\dot{\varphi}^4\theta^2\hat{\triangledown}_\mu\dot{\varphi} + \frac{5}{9}\dot{\varphi}^4\theta^3\hat{\triangledown}_\mu\varphi - \frac{2}{3}\dot{\varphi}^5\theta\hat{\triangledown}_\mu\theta + \dot{\varphi}^5\theta\hat{\triangledown}^\alpha\sigma_{\mu\alpha} + \dot{\varphi}^5\theta\hat{\triangledown}^\alpha{\varpi}_{\mu\alpha} \right],\\
\label{perturbed4}
\pi_{\mu\nu}^\varphi &\doteq& \frac{c_4}{M^6}\left[  -\dot{\varphi}^4 \left( \dot{\sigma}_{\mu\nu} - \hat{\triangledown}_{\langle\mu}A_{\nu\rangle} - \mathcal{E}_{\mu\nu}\right) - \left( 6\ddot{\varphi}\dot{\varphi}^2 + \frac{2}{3}\dot{\varphi}^3\theta\right)\hat{\triangledown}_{\langle\mu}\hat{\triangledown}_{\nu\rangle}\varphi - \left(6\ddot{\varphi}\dot{\varphi}^3 + \frac{4}{3}\dot{\varphi}^4\theta\right)\sigma_{\mu\nu}\right] \nonumber \\
&& + \frac{c_5}{M^9}\left[ -\left(\dot{\varphi}^5\dot{\theta} +\dot{\varphi}^5\theta^2 + 6\ddot{\varphi}\dot{\varphi}^4\theta\right)\sigma_{\mu\nu} -  \left(\dot{\varphi}^5\theta + 3\ddot{\varphi}\dot{\varphi}^4\right)\dot{\sigma}_{\mu\nu} -  \left( 4\ddot{\varphi}\dot{\varphi}^3\theta + \dot{\varphi}^4\dot{\theta} + \frac{1}{3}\dot{\varphi}^4\theta^2 \right)\hat{\triangledown}_{\langle\mu}\hat{\triangledown}_{\nu\rangle}\varphi \right. \nonumber \\
&& \left. \ \ \ \ \ \ \ \ \ \ \ \ \ + \left(\dot{\varphi}^5\theta + 3\ddot{\varphi}\dot{\varphi}^4 \right)\hat{\triangledown}_{\langle\mu}A_{\nu\rangle} - 6\ddot{\varphi}\dot{\varphi}^4\mathcal{E}_{\mu\nu} \right],
\eq
\end{widetext}
in which $\hat{\Box}\equiv\hat{\nabla}^\mu\hat{\nabla}_\mu$. In Eq.~(\ref{perturbed4}), $\hat{\nabla}_{\langle\mu}\hat{\nabla}_{\nu\rangle}\varphi$ and $\mathcal{E}_{\mu\nu} = u^{\alpha}u^{\beta}\mathcal{W}_{\mu\alpha\nu\beta}$ are PSTF rank-2 tensors, both of which live in the 3-dimensional hypersurface perpendicular to $u^\mu$ ($u^\mu\hat{\nabla}_{\langle\mu}\hat{\nabla}_{\nu\rangle}\varphi  = u^\mu\mathcal{E}_{\mu\nu} = 0$), where $\mathcal{W}_{\mu\alpha\nu\beta}$ is the Weyl curvature tensor. The equality symbol $\doteq$ means that we have neglected the terms which are higher than linear order in small perturbations. Angular brackets indicate trace-free quantities.

Note that the $\mathcal{L}_2$ and $\mathcal{L}_3$ terms do not contribute to the Galileon field anisotropic stress.

The Galileon field EOM, again, up to linear order in small perturbations, is given by
\begin{widetext}
\begin{eqnarray}
\label{perturbed EoM}
0 &\doteq& c_2\left[\ddot{\varphi} + \hat{\Box}\varphi + \dot{\varphi}\theta \right] + \frac{c_3}{M^3} \left[ 4\ddot{\varphi}\dot{\varphi}\theta  + \frac{8}{3}\dot{\varphi}\theta\hat{\Box}\varphi + 4\ddot{\varphi}\hat{\Box}\varphi + 2\dot{\varphi}^2\theta^2 + 2\dot{\varphi}^2\dot{\theta} - 2\dot{\varphi}^2 \hat{\nabla} \cdot A\right] \nonumber \\
&&  + \frac{c_4}{M^6}\left[  6\ddot{\varphi}\dot{\varphi}^2\theta^2 + 4\dot{\varphi}^3\dot{\theta}\theta + 2\dot{\varphi}^3\theta^3 + 8\ddot{\varphi}\dot{\varphi}\theta\hat{\Box}\varphi + \frac{26}{9}\dot{\varphi}^2\theta^2\hat{\Box}\varphi - 4\dot{\varphi}^3\theta\hat{\nabla}\cdot A + 4\dot{\varphi}^2\dot{\theta}\hat{\Box}\varphi + 3\ddot{\varphi}\dot{\varphi}^2\hat{R} + \frac{1}{3}\dot{\varphi}^3\theta\hat{R}\right] \nonumber \\
&& + \frac{c_5}{M^9}\left[ \frac{5}{9}\dot{\varphi}^4\theta^4   +\frac{20}{9}\ddot{\varphi}\dot{\varphi}^3\theta^3  +\frac{5}{3}\dot{\varphi}^4\dot{\theta}\theta^2 +\frac{8}{9}\dot{\varphi}^3\theta^3\hat{\Box}\varphi +\frac{1}{2}\dot{\varphi}^4\dot{\theta}\hat{R} \right. \nonumber \\
&&\ \ \ \ \ \ \ \ \ \ \ \ \ \ \ \ \ \  \left. +\frac{1}{6}\dot{\varphi}^4\theta^2\hat{R} - \frac{5}{3}\dot{\varphi}^4\theta^2\hat{\nabla} \cdot A  + 4\ddot{\varphi}\dot{\varphi}^2\theta^2\hat{\Box}\varphi + \frac{8}{3}\dot{\varphi}^3\dot{\theta}\theta\hat{\Box}\varphi + 2\ddot{\varphi}\dot{\varphi}^3\theta\hat{R} \right].
\end{eqnarray}
\end{widetext}

Eqs.~(\ref{perturbed1} - \ref{perturbed EoM}) also contain the background (zeroth-order terms) as the non-hatted terms. Here, for clarity of exposition and because these terms enter the physical calculations separately, we extract the background expression for the Galileon field density, pressure and equation of motion, which follow, respectively, as:

\bq\label{density-background}
\bar{\rho}_\varphi &=& c_2\left[ \frac{1}{2}\dot{\varphi}^2\right] + \frac{c_3}{M^3} \left[ 2\dot{\varphi}^3\theta\right]  +\frac{c_4}{M^6}\left[ \frac{5}{2}\dot{\varphi}^4\theta^2\right]     \nonumber \\
&& + \frac{c_5}{M^9}\left[\frac{7}{9}\dot{\varphi}^5\theta^3\right],
\eq

\bq\label{pressure-background}
\bar{p}_\varphi &=& c_2\left[ \frac{1}{2}\dot{\varphi}^2\right] + \frac{c_3}{M^3} \left[ -2\ddot{\varphi}\dot{\varphi}^2\right] \nonumber \\
&& + \frac{c_4}{M^6}\left[ -4\ddot{\varphi}\dot{\varphi}^3\theta - \dot{\varphi}^4\dot{\theta} - \frac{1}{2}\dot{\varphi}^4\theta^2\right] \nonumber \\
&& + \frac{c_5}{M^9}\left[-\frac{5}{3}\ddot{\varphi}\dot{\varphi}^4\theta^2 - \frac{2}{3}\dot{\varphi}^5\dot{\theta}\theta - \frac{2}{9}\dot{\varphi}^5\theta^3\right],
\eq

\bq\label{background-EoM}
0 &=& c_2\left[\ddot{\varphi} + \dot{\varphi}\theta \right] + \frac{c_3}{M^3} \left[ 4\ddot{\varphi}\dot{\varphi}\theta + 2\dot{\varphi}^2\theta^2 + 2\dot{\varphi}^2\dot{\theta}\right] \nonumber \\
&&  + \frac{c_4}{M^6}\left[  6\ddot{\varphi}\dot{\varphi}^2\theta^2 + 4\dot{\varphi}^3\dot{\theta}\theta + 2\dot{\varphi}^3\theta^3 \right] \nonumber \\
&& + \frac{c_5}{M^9}\left[ \frac{5}{9}\dot{\varphi}^4\theta^4   +\frac{20}{9}\ddot{\varphi}\dot{\varphi}^3\theta^3  +\frac{5}{3}\dot{\varphi}^4\dot{\theta}\theta^2 \right],
\eq
where we have neglected the overbars on $\varphi$ to lighten the notation. The heat flux $q_\mu^\varphi$ and anisotropic stress $\pi_{\mu\nu}^\varphi$ contain only perturbed quantities and vanish at the cosmological background level. Note that assuming the de Sitter limit where $\theta$ and $\dot{\varphi}$ are both constants, it can be confirmed that Eqs.~(\ref{density-background}) and (\ref{pressure-background}) satisfy $\rho_\varphi+P_\varphi=0$, that is $w = -1$.

Eqs.~(\ref{background1}, \ref{background2}, \ref{radiation-matter-density}, \ref{radiation-matter-pressure}, \ref{density-background}, \ref{pressure-background}, \ref{background-EoM}) form a set of differential equations for the background evolution of the Universe, whereas Eqs.~(\ref{perturbed1} -- \ref{perturbed EoM}) are the expressions which enter the equations in {\tt CAMB} to solve for the evolution of the linear perturbations \cite{Barreira:2012kk, camb_notes}. 

\subsection{Cosmology of the Galileon model}

The cosmological features of the Galileon field have been studied in the literature in great detail both at the background \cite{PhysRevD.82.103015, PhysRevD.80.024037, DeFelice:2010pv, PhysRevD.82.124054, Gannouji:2010au} and at the linear perturbation level \cite{Barreira:2012kk, DeFelice:2010as, Appleby:2011aa, Neveu:2013mfa}. Here we just briefly outline the most distinctive aspects of the covariant Galileon model.

\subsubsection{Background}

In \cite{DeFelice:2010pv} the authors have shown the existence of a tracker solution of the dynamical background equations which approaches de Sitter evolution ($w = -1$) at the present time. This happens after periods of radiation and matter domination, thus allowing for a viable expansion history. The tracker is characterized by phantom evolution ($w < -1$) of the Galileon field in the past and works as an attractor for other solutions with different initial conditions. In \cite{PhysRevD.82.124054} it was shown that the background cosmological data prefers solutions which approach the tracker at late times so that the phantom period of the evolution only occurs close to today and the equation of state parameter $w$ does not become too much smaller than $w = -1$.

A noteworthy aspect here is the possibility of having ghost-free phantom dynamics \cite{Deffayet:2010qz, Pujolas:2011he, Sawicki:2012re} which could lead to clear signatures in the expansion rate of the Universe \cite{Zhao:2012aw}, thus helping to distinguish this model from other models such as $\Lambda$CDM or Quintessence.  

\subsubsection{Linear perturbations}

The full linear perturbation equations of the Galileon model have been derived in \cite{DeFelice:2010as, Barreira:2012kk}, where it was also shown that the quasi-static limit (the limit where the time derivatives are neglected relative to spatial derivatives) is in general a good approximation for the full equations. In \cite{DeFelice:2011hq} the authors derived the linear perturbation equations for the full Horndeski Lagrangian, which includes the Galileon model studied here.

In \cite{Barreira:2012kk} we showed that the Galileon field can cluster strongly at the linear level since early times and in a way comparable to the clustering of matter. Such a strong clustering signal can be attributed to the non-trivial behavior of the pressure perturbation and the anisotropic stress in the Galileon model. 

The clustering of the Galileon field is intimately related with the time evolution of the gravitational potential. In the Galileon model, the gravitational potential can evolve significantly even deep inside the matter dominated era (in contrast to the $\Lambda$CDM paradigm, where it remains constant throughout matter domination). This leads to very clear signatures in the low-$l$ region of the CMB temperature power spectrum where the ISW effect, precisely determined by the time evolution of the gravitational potential, makes the dominant contribution. In addition to a nonstandard time variation, the gravitational potential can experience an overall deepening with respect to the $\Lambda$CDM prediction, which also leaves distinct imprints on certain cosmological observables. In particular, it can strongly modify the growth rate of matter density fluctuations \cite{Appleby:2012ba, Okada:2012mn, Barreira:2012kk, Neveu:2013mfa} and also make very different predictions for observables such as weak lensing and the cross correlation between the ISW effect and the galaxy distribution \cite{Barreira:2012kk}.

\section{Methodology}\label{Methodology}

\subsection{The Galileon cosmological parameter space}

We restrict ourselves to a flat geometry for the background space-time. In this case the full cosmological parameter space we consider is eleven-dimensional with six cosmological parameters: $\Omega_{c}^0$, $\Omega_{b}^0$, $h$, $\tau$, $n_s$ and $\rm{log}[10^{10}A_s]$, which are, respectively, the present-day values of the fractional densities of dark matter and baryonic matter, the Hubble expansion rate $h = H_0/(100~\rm{km/s/Mpc})$, the optical depth to reionization, the scalar spectral index ($n_s$) and amplitude ($A_s$) of the primordial power spectrum of the scalar fluctuations; plus the five Galilean parameters $c_2$, $c_3$, $c_4$, $c_5$ and $\dot{\bar{\varphi}}_i$, where $\dot{\bar{\varphi}}_i$ is background value of the time derivative of the Galileon field at the initial time $t_i$ (or redshift $z_i$) when the calculation starts. In this paper we take $z_i = 10^6$. The energy density of radiation is held fixed at $\Omega_{r0}{h}^2 = 4.18\times10^{-5}$, which is given by the temperature of the CMB photons, $T_{\rm{CMB}} = 2.725 \rm{K}$, with an effective number of massless neutrinos given by $N_{\rm{eff}}^{\rm{massless\ \nu}} = 3.04$ \cite{Hinshaw:2012fq}.

Since we are assuming a spatially flat background Universe, we require that the fractional energy densities today satisfy $\Omega_{\varphi}^0 + \Omega_{c}^0 + \Omega_{b}^0 + \Omega_{r}^0 = 1$. This constraint equation means that one of the Galileon parameters $c_{2-5}$, $\dot{\bar{\varphi}}_i$ can be uniquely derived from the others. In this work, we choose $c_2$ as the derived parameter, and follow a trial-and-error approach to find the value of $c_2$ for which the universe is spatially flat. Note that for each new trial value of $c_2$, we need to evolve the whole set of background equations from $z=z_i$ to $z=0$, to check if the constraint is satisfied. This procedure can take a significant time, particularly because we have to sample many points in the parameter space during the MCMC search. To speed up this process, we designed an algorithm that takes adaptive step lengths in setting the trial value of $c_2$, as well as adopting a conservative criterion by stopping the trial-and-error search when $|\Omega_{\varphi}^0 + \Omega_{c}^0 + \Omega_{b}^0 + \Omega_{r}^0-1|<10^{-3}$. This accuracy is sufficient for all our numerical results. Because $c_2$ is now a derived parameter, the dimensionality of the parameter space is reduced by one.

The value of the Galileon background density $\bar{\rho}_{\varphi,i}$, at the starting redshift, is determined by the values of $\dot{\bar{\varphi}}_i$ and $\theta_i$, the latter being given by the fixed matter and radiation components via Eqs.~(\ref{background1}, \ref{radiation-matter-density}) (the Galileon energy density is negligible at the very early times). Because this is a more physically meaningful parameter than $\dot{\bar{\varphi}}_i$, despite sampling through different values of $\dot{\bar{\varphi}}_i$ in the MCMC search we use the ratio between the Galileon and matter energy densities at the starting redshift, $\bar{\rho}_{\varphi,i} / \bar{\rho}_{m,i}$, when we quote the observational constraints below. Moreover, Eq.~(\ref{density-background}) can be used to translate in between the values of $\bar{\rho}_{\varphi,i}$ and $\dot{\bar{\varphi}}_i$ via $\bar{\rho}_{\varphi,i} \approx 7c_5 \dot{\bar{\varphi}}_i^5\theta_i^3 / (9M^9)$. This approximation holds at sufficiently early times where the Galileon density is negligible\footnote{Otherwise, one would need to take the initial Galilean density into account in the determination of $\theta_i$, which is then given, approximately, by $\theta_i \approx \sqrt{3\kappa(\rho_{mi} + \rho_{ri})}$, via Eq.~(\ref{background1}).} provided also that $c_5$ is not too much smaller than the values of the other $c_n$ (which is the case for the best-fitting regions of the parameter space).

To solve for the evolution of the linear perturbations, we also need the initial conditions of the Galileon field perturbation and its time derivative at $z_i$, which are, in principle, also free parameters. However, we know that these numbers must be very small at high redshift. As a result, we have set both of them to be exactly zero, and checked that our results are insensitive to sufficiently small changes around these initial values for all length scales (or $k$-modes) that are of interest to us. Here, `sufficiently small' means small enough to still be in the regime of linear perturbation theory. Typically, such a condition is quoted as $\delta\varphi_i \ll \bar{\varphi}_i$. However, the background value of the field is irrelevant in the covariant Galileon model since the background equations only involve $\dot{\bar{\varphi}}$ and $\ddot{\bar{\varphi}}$ and not $\bar{\varphi}$. As an alternative, we adopt $\delta\varphi_i \ll \dot{\bar{\varphi}}_i / H_i$ as a criterion for the validity of linear perturbation analysis, and this restricts $\delta\varphi_i$ and $\dot{\delta\varphi}_i$ to be so small that it makes no practical difference if they are set to be exactly zero.

\subsection{Scaling degeneracy in the Galileon model}

By looking at Eqs.~(\ref{perturbed1} - \ref{perturbed EoM}), which are the equations that fully govern the Galileon physics, we note that they are invariant under the following transformations:
\bq\label{scaling relation}
c_2 &\longrightarrow& c_2' = c_2 / B^2, \nonumber \\
c_3 &\longrightarrow& c_3' = c_3 / B^3, \nonumber \\
c_4 &\longrightarrow& c_4' = c_4 / B^4, \nonumber \\
c_5 &\longrightarrow& c_5' = c_5 / B^5, \nonumber \\
\varphi &\longrightarrow& \varphi' = \varphi  B,
\eq
in which $B$ is an arbitrary constant and the transformation of $\varphi$ holds for both the background and perturbation parts.This scaling relation is what allows \cite{DeFelice:2010pv} and \cite{Neveu:2013mfa} to use different sets of parameters to characterize the Galileon model. The reason for this scaling lies in the fact that each of the Galileon Lagrangians $\mathcal{L}_i$ yields terms which all have the same power in the Galileon field $\varphi$ (for the counting of the power, the time and spatial derivatives of $\varphi$ are treated equally as $\varphi$). Note that the transformation of Eqs.~(\ref{scaling relation}) preserves the signs of the parameters $c_2$ and $c_4$, but not those of $c_3$, $c_5$ and $\dot{\bar{\varphi}}_i$ (if $B < 0$). This illustrates that the scaling also contains a `reflection symmetry' in the $c_3 - c_5 - \dot{\bar{\varphi}}_i$ subspace, which is associated with the sign of $B$, instead of its magnitude.

The existence of such a scaling relation is non-trivial in dark energy and modified gravity models. For example, the EOM of a Quintessence field $\varphi$ with a self-interacting potential $V(\varphi)$ is $\ddot{\varphi} + \theta\dot{\varphi} + {\rm d}V(\varphi)/{\rm d}\varphi = 0$; therefore, unless $V(\varphi) \propto \varphi^2$, it is impossible to allow for this scaling relation. From a practical point of view, the scaling of the Galileon field $\varphi$ is realized by rescaling its time derivative at the initial time, $\dot{\bar{\varphi}}_i$ \footnote{In principle, to achieve the exact scaling at the linear perturbation level, one has to resize the Galileon field perturbation $\delta\varphi_i$ and its time derivative $\dot{\delta\varphi}_i$, accordingly. However, thanks to our choices of initial conditions, namely $\delta\varphi_i=\dot{\delta\varphi}_i=0$, such resizing does not need to be done explicitly.}. As a result, according to Eqs.~(\ref{scaling relation}), the impact of smaller values of $\dot{\bar{\varphi}}_i$ can always be compensated by larger values of the $c_n$ parameters and vice versa, thus making the $c_n$ parameters unbounded and preventing a proper constraint of the parameter space.

The scaling relation allows one to further reduce the dimensionality of the Galileon subspace of parameters by one. This can be done by using one of the Galileon parameters as a reference to write down invariant quantities under the scaling (e.g., in \cite{Neveu:2013mfa}, the reference parameter is the present-day value of the Galileon field time derivative). In this paper, we take $c_3$ as the reference parameter and therefore the invariant quantities are
\bq\label{scale-invariant}
\left(\frac{c_2}{c_3^{2/3}}, \frac{c_4}{c_3^{4/3}}, \frac{c_5}{c_3^{5/3}}, c_3^{1/3}\dot{\bar{\varphi}}_i\right).
\eq
Note that $\bar{\rho}_{\varphi, i} / \bar{\rho}_{m,i}$ is, by definition, invariant under the scaling (c.f.~Eq.~(\ref{perturbed1})). When running the Markov chains, while one could allow for all of the parameters to vary and apply the constraints to the set of Eq.~(\ref{scale-invariant}), it is easier to fix the value of $c_3$. This will increase the convergence rate of the MCMC algorithm since now there is one fewer dimension to sample from and the unboundedness of the $c_n$ parameters can be avoided.


Note that in principle any other Galilean parameter could be used as the reference to write down the scaling invariant terms. In practice, however, $c_2$ cannot be used as the fixed parameter since it is the parameter which we tune to yield a consistent cosmological background evolution with the required amount of dark energy to make the Universe spatially flat, and is unknown {\it a priori}. Practically, it is not a good idea to fix $c_4$ or $c_5$ either, because when all parameters are unfixed they are more likely to cross or become very close to zero -- we have checked that, for example, fixing $c_5=1$ (which is roughly the best-fitting value of $c_5$ when $c_3=10$) causes $c_3$ and $c_4$ to become essentially unbounded since $c_5$ can be as small as $10^{-5}$ when $c_3=10$. However, having derived the constraints with $c_3$ fixed, one can always use the scaling of Eqs.~(\ref{scaling relation}) to scale the constrained regions to the case where, for example, $c_2 = -1$, which, in our convention in Eq.~(\ref{Galileon action}), corresponds to the Galileon field having a standard scalar kinetic term, but with a different sign.

In summary, the nine-dimensional parameter space we aim to constrain in this paper is specified by
\begin{eqnarray}
\left( \Omega_{c0}, \Omega_{b0}, h, \tau, n_s, {\log}\left[10^{10}A_s\right], \frac{c_4}{c_3^{4/3}}, \frac{c_5}{c_3^{5/3}}, c_3^{1/3}\dot{\bar{\varphi}}_i \right),\nonumber
\end{eqnarray}
in which the first six are cosmological parameters and the last three are the Galileon (physical) parameters. The remaining Galileon parameters, namely $c_2c_3^{-2/3}$ and $\bar{\rho}_{\varphi,i}/\bar{\rho}_{m,i}$, are treated as derived parameters.
\subsection{Theoretical constraints}

In addition to the observational constraints that we will outline in the next section, the Galilean parameter space is {\it a priori} constrained by the requirement to avoid the appearance of theoretical instabilities. The Galileon Lagrangian, being a subset of the more general Horndeski Lagrangian, is automatically protected against the propagation of Ostrogradsky ghosts as the equations are retained up to second order \cite{Woodard:2006nt}. However, other sorts of theoretical pathologies may still arise. 

In this paper we consider the conditions for each point in parameter space not to develop ghost degrees of freedom or Laplace instabilities in the scalar sector of the linear perturbations (see e.g. \cite{DeFelice:2010nf, DeFelice:2011bh, Appleby:2011aa} for a discussion and derivation of the stability conditions). The no-ghost and no-Laplace stability conditions, despite applying to the scalar perturbations, depend only on background quantities \cite{Appleby:2011aa}. When the MCMC algorithm tries a new point in parameter space, our code first solves the background evolution, testing whether or not it satisfies all the stability criteria. The calculation of the evolution of the perturbations and the subsequent likelihood evaluation is only performed if the point is theoretically viable. We only test the theoretical stability of any given point in the past, since there is no evidence that the instabilities cannot develop in the unprobed future.

One could also consider other theoretical conditions such as those which ensure that the Galileon field perturbation does not propagate superluminally, i.e. $c_s^2>1$ (see e.g.~\cite{deFromont:2013iwa}). However, such cases do not necessarily imply the existence of pathologies such as the violation of causality (see e.g. \cite{Babichev:2007dw, Burrage:2011cr}) and therefore we do not employ them. We also do not rule out {\it a priori} cases where $\bar{\rho}_\varphi < 0$ at some point in time, but instead let the data decide their viability.

We want to stress that the theoretical constraints are a convenient way to select only those points which give viable perturbation evolution, and once these constraints are satisfied so that a trial parameter point is not rejected straightaway, they play no further role in the calculation of likelihoods.

\subsection{Datasets}

To derive the constraints on the Galileon parameter space we use the CMB data of the full WMAP 9-year final release \cite{Hinshaw:2012fq} in the form of the {\tt FORTRAN} likelihood software provided by the WMAP team, for which we have adapted {\tt CosmoMC}. Despite being WMAP's final release, the dataset will soon be replaced by the upcoming Planck satellite data. However, we expect that the differences between these two datasets will be of little importance as regards the Galileon model constraints because, as we will see in the following sections, the major impact of the Galileon model on the CMB happens on the largest angular scales, where Planck is not expected to perform much better than WMAP due to the limits imposed by cosmic variance. Note that unlike \cite{Nesseris:2010pc, Appleby:2012ba, Neveu:2013mfa}, we do not use the WMAP distance priors, which are derived parameters and require assumptions to be made about the background cosmology. Moreover, below we will see how the Galileon model can have an impact on the largest angular scales of the CMB and also on the amplitude of the acoustic peaks, which illustrates the advantage of using the full CMB data over using only the information encoded in the positions of the acoustic peaks. 

As a low redshift probe, we consider the three year sample of the Supernova Legacy Survey (SNLS) project \cite{Guy:2010bc} which contains 472 type Ia supernovae ranging from $z \approx 0.15$ to $z \approx 1.1$. When quoting the results, we marginalize over the parameters that are used in the calibration of the intrinsic luminosity of the SNIa of the SNLS sample.

To complement the constraints from SNIa luminosity distances, we also use measurements of the angular scale of the BAO feature in the galaxy distribution whose dependence on the background expansion rate differs from the SNIa constraints and, therefore, probes a different region of the parameter space. We use the BAO measurements from the 6df Galaxy Survey at redshift $z = 0.106$ \cite{Beutler:2011hx}, from the Sloan Digital Sky Survey (SDSS) DR7 at $z = 0.275$ and $z = 0.35$ \cite{Percival:2009xn} and from the SDSS-III Baryon Oscillation Spectroscopic Survey (BOSS) at $z = 0.57$ \cite{Anderson:2012sa}. 

\subsubsection{Why we do not use growth rate and clustering data}

Several studies have shown that the modifications of gravity in the Galileon model can significantly enhance the growth of linear matter fluctuations on sub-horizon scales \cite{Barreira:2012kk, DeFelice:2010as, Appleby:2011aa, Okada:2012mn, Appleby:2012ba, Neveu:2013mfa}. In particular, in \cite{Okada:2012mn, Appleby:2012ba, Neveu:2013mfa}, the authors used measurements of the linear growth factor to place strong constraints on the Galileon model. These studies have shown that there is an unavoidable tension in the Galileon model between its ability to fit, at the same time, background and growth data. However, these conclusions assume the validity of linear perturbation theory in the Galileon model on the scales where the growth factor is measured. In GR, one would expect the scales probed by current growth factor measurements to be mildy nonlinear \cite{Jennings:2011}. However, for the Galileon model we do not yet have reliable knowledge about the true scale on which the Vainshtein screening starts to be important and how it affects structure formation: as this is a purely nonlinear effect, it is by definition not present in a linear theory analysis. For example, numerical simulations have shown that in other modified gravity models such as the $f(R)$ and dilaton \cite{Brax:2010gi, Brax:2011ja, Brax:2012nk}, linear perturbation theory can fail even on scales as large as $k \sim 0.01~{h}\rm{Mpc}^{-1}$ because of the chameleon screening \cite{Jennings:2012pt, Brax:2012nk, Li:2012by}. It is reasonable to suspect that similar situations might occur in the Galileon model.

Our modifications to the {\tt CAMB} code allow us to obtain the Galileon predictions for the linear matter power spectrum and growth rate \cite{Barreira:2012kk} and we could in principle use these to place further constraints on the model. However, given the above reasoning we remain cautious about using clustering data for now. We argue that a better understanding of the true impact of the Vainshtein screening is needed before attempting a more rigorous confrontation of the predicted clustering power and growth rate with the observational data. Such an analysis of the Vainshtein effect in the Galileon model will be left for future work (see also \cite{Bellini:2012qn, Kimura:2011dc, Babichev:2011iz, DeFelice:2011th, Hiramatsu:2012xj, Burrage:2010rs, Iorio:2012pv, Brax:2011sv, deRham:2012fw, deRham:2012fg} for some existing studies in this direction). In the present paper, we limit ourselves to what the CMB and background data can tell about the parameter space of the Galileon gravity model.

\section{Results}\label{Results}

\begin{figure*}
	\centering
	\includegraphics[scale=0.39]{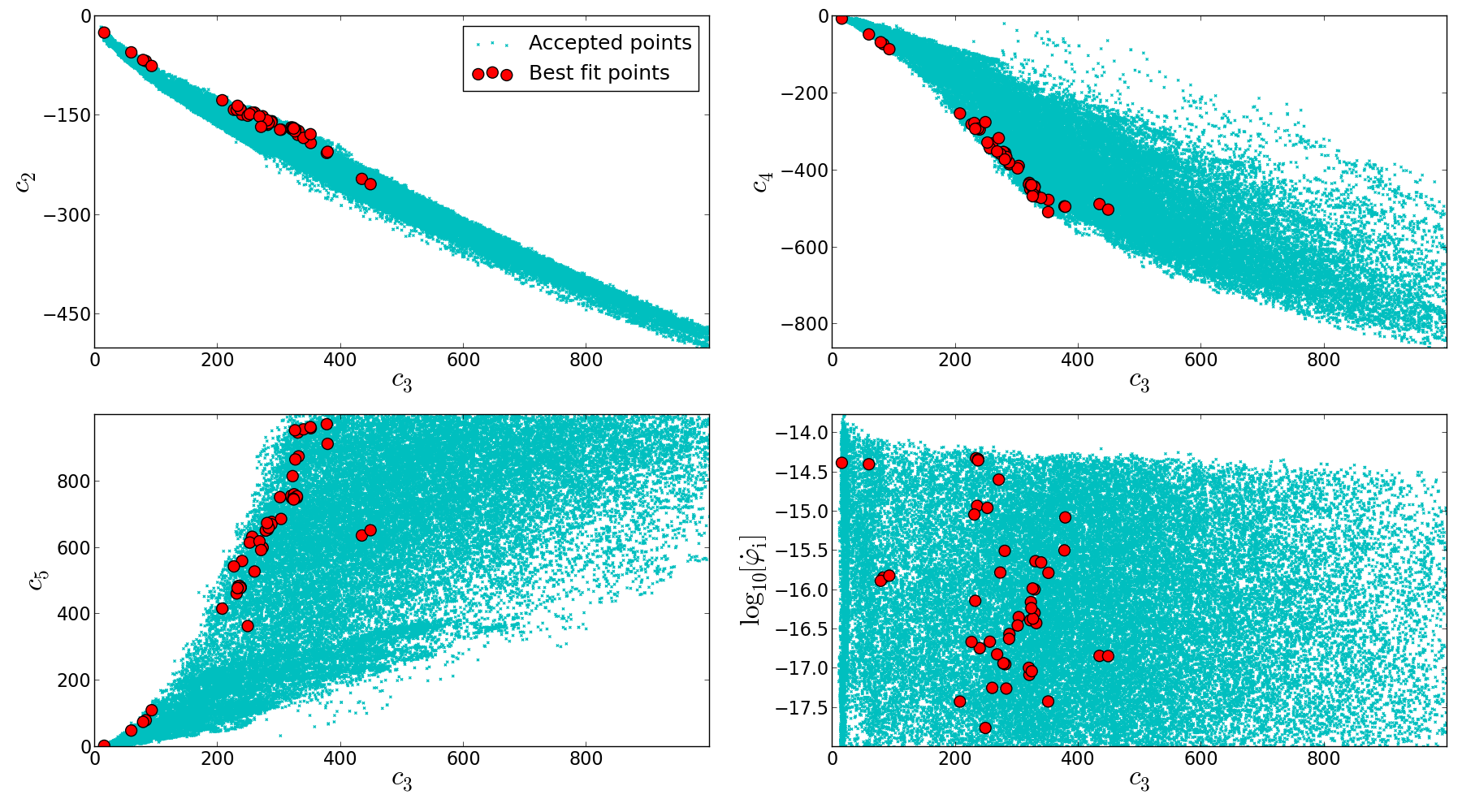}
	\caption{(Color online) Points accepted by the Metropolis-Hastings algorithm, including those sampled during the burn-in period, using WMAP9+SNLS+BAO (blue dots). The points are projected onto different planes of the Galileon subset of the parameter space. All of the Galileon parameters were allowed to vary in order to manifest the scaling degeneracy. The remaining cosmological parameters were also sampled (not shown; see Fig.~\ref{cosmological-constraints}). There are approximately $48000$ points (around $8000$ per chain) and the large red circles represent the fifty best-fitting points.}
\label{scaling-proof}\end{figure*}

\subsection{Numerical details}

We use the publicly available {\tt CosmoMC} code \cite{Lewis:2002ah} to carry out the formal exploration of the full Galileon cosmological parameter space, using the Metropolis-Hastings algorithm (see \cite{Lewis:2002ah} for a concise description) as the MCMC method to draw the samples from the posterior probability distribution, which is determined by the observational data. The code is compatible with our modified version of the publicly available {\tt CAMB} code \cite{camb_notes}, which solves for the background and perturbation evolution in Galileon models \cite{Barreira:2012kk}. The results we show in this section were obtained by running six chains in parallel with the stoping convergence criterion $R - 1 < 0.02$, where $R$ is the Gelman and Rubin statistic given by $R = $"variance of chains means"/"mean of chains variances" \cite{An98stephenbrooks}. The MCMC algorithm is used to explore the parameter space with linear steps along all directions, except the $\dot{\bar{\varphi}}_i$ direction, where the steps are logaritmic since this parameter can take values that differ by a few orders of magnitude. The estimation of the likelihood from the samples was performed using the routines in the {\tt Getdist} software supplied in the {\tt CosmoMC} package. We do not consider the first half of the chains in the likelihood evaluation to eliminate points sampled during the `burn-in' period of the chains. When plotting the marginalized parameter distributions and the respective probability limit contours, we smooth the likelihood surface by applying a Gaussian filter (in addition to the smoothing already performed by the {\tt Getdist} routines). We made sure that the smoothing does not affect our results. In Table \ref{table-mean}, the results are for the unsmoothed distributions as determined by the {\tt Getdist} evaluation. We have also made sure that the prior range limits specified for each of the parameters were sufficiently far away from where the likelihood distribution is non-negligible.

In addition to the standard modifications needed to link {\tt CAMB} with {\tt CosmoMC} when new parameters are added or alternative cosmological models are studied, we have also changed the way in which the ${h}$ cosmological parameter is sampled. In the latest version of the code, the MCMC algorithm samples over the parameter $r_s^{\rm rec} / d_A^{\rm rec}$ (known in the code as the parameter {\tt theta}) where $r_S^{\rm rec}$ and $d_A^{\rm rec}$ are, respectively, the sound horizon at recombination and the angular diameter distance to recombination. Having sampled $r_s^{\rm rec} / d_A^{\rm rec}$, the code then finds the value of $h$ by a trial-and-error procedure. The reason why it is done this way is because $r_s^{\rm rec} / d_A^{\rm rec}$ is much less correlated than ${h}$ with the rest of the cosmological parameters, thus improving the performance of the parameter space exploration. However, in our case one has also the trial-and-error search for $c_2$ which can be in conflict with the search for $h$. The value of $c_2$ is fixed when the code is trying values of ${h}$, and in the Galileon model this can very easily lead to several numerical problems related with cases where the background evolution could develop `fake' ghosts or Laplace instabilities \footnote{`Fake' here is in the sense that such cases could have been free of instabilities if $c_2$ took the correct value for the consistent background evolution. However, this can only be done by simultaneously searching for $c_2$ and $h$ using trial-and-error and it is, therefore, much more time consuming.}. Consequently, although one could in principle work its way around these problems, for simplicity, we choose to sample ${h}$ directly, even if this happens at the cost of having slightly slower runs.

The trial-and-error search for $c_2$ can be very time consuming and to speed it up we have adopted adaptive step lengths in setting trial values of $c_2$. If the step length becomes very large, it is possible that the trial value of $c_2$ becomes too far away from the right value so that the numerical solver of the EOM fails. However, we have checked that this is very rare with our algorithm, and by comparing with the results from fixing the step length we have confirmed that it does not affect our numerical constraints and conclusions.

\subsection{Scaling degeneracy}

As discussed in the previous section, one of the Galileon parameters should be fixed when running the chains to break the scaling degeneracy of the Galileon equations. However, in order to gain some insight into how the scaling degeneracy manifests itself in the parameter space, we have first run a set of chains where all of the parameters are free to vary. The result is shown in Fig.~\ref{scaling-proof}, where we plot all the points accepted by the MCMC algorithm (cyan dots) for four different planes of the Galileon parameter subspace for chains constrained using the combined WMAP9+SNLS+BAO dataset. There are approximately $48000$ points ($8000$ per chain). We highlight the points with the fifty highest likelihood values (red circles). The rest of the cosmological parameters were allowed to vary as well (not shown in this plot, see Fig.~\ref{cosmological-constraints} below). The parameters $c_3$, $c_4$ and $c_5$ were sampled from the interval $\left[ -1000, 1000\right]$ to prevent the chains from spending too much time searching larger and larger values of the $c_n$. The starting point of the $c_n$ for each of the chains was set to be sufficiently close to zero so that the algorithm could quickly select the signs for the parameters which best fit the data. For practical reasons, we sample only positive (or only negative) values of $\dot{\bar{\varphi}}_i$, since its absolute value can be very small. We point out, however, that this does not mean we are rulling out the regions of the parameter space where $\dot{\bar{\varphi}}_i < 0$, since such regions can always be found by simultaneously fliping the signs of $c_3$, $c_5$ and $\dot{\bar{\varphi}}_i$, as allowed by the scaling transformations of Eq.~(\ref{scaling relation}).

As expected from the scaling relations of Eqs.~(\ref{scaling relation}), one finds a long and narrow region of degeneracy in the Galileon subset of the parameter space, along which the likelihood is kept constant. Note that although it seems that the best-fitting points are confined to  $c_2 \gtrsim -200$, $c_3 \lesssim 400$ and $c_4 \gtrsim -600$, this happens only because the points have reached the prior range limit of $1000$ in the $c_5$ direction, which therefore `artificially' constrains the other parameters. We have checked that the degeneracy region keeps increasing on increasing the size of the prior ranges. The difference in $\chi^2 = -2{\log} P$ (where $P$ is the posterior) between the best-fitting and the fiftieth best-fitting points is $\Delta\chi^2_{1th, 50th} \sim -1$, but the likelihood does not change monotonically along any direction of the parameter space. The result shown in Fig.~\ref{scaling-proof} is in partial disagreement with the conclusions drawn in \cite{Appleby:2012ba}. In the latter, the authors found a long and narrow region of degeneracy along which the likelihood decreases for values of $c_n$ much larger or much smaller than unity. We agree that the long region of degeneracy exists. However, the likelihood does not change appreciably towards larger values of the parameters $c_n$, which is what one would expect in light of the scaling relation described by Eqs.~(\ref{scaling relation}).

We see that the best-fitting points all lie in the region of parameter space where $c_2 < 0$, $c_3 > 0$, $c_4 < 0$, $c_5 > 0$ when $\dot{\bar{\varphi}}_i > 0$. This means that the sign of the fixed parameter $c_3$ should be the same as $\dot{\bar{\varphi}}_i$. For instance, if we were to fix a negative value of $c_3$ while sampling only positive values of $\dot{\bar{\varphi}}_i$, we would be discarding, {\it a priori}, the portion of the parameter space that contains the best-fittings points (red dots). Recall that the reflection symmetry only holds if the three parameters $c_3$, $c_5$ and $\dot{\bar{\varphi}}_i$ all flip their signs. Throughout, the results will refer to sets of chains where $c_3 = 10$ and $\dot{\bar{\varphi}}_i > 0$, and we shall quote the final constraints in terms of the invariant combinations of Eq.~(\ref{scale-invariant}). We stress that these combinations are also invariant under the simultaneous change of the sign of the $c_3$, $c_5$ and $\dot{\bar{\varphi}}_i$ parameters. In what follows we will refer to $c_p$ and $c_p / c_3^{p/3}$ interchangeably, where $p = 2,4,5$.

\subsection{Parameter space constraints}\label{constraints}

\begin{figure*}
	\centering
	\includegraphics[scale=0.39]{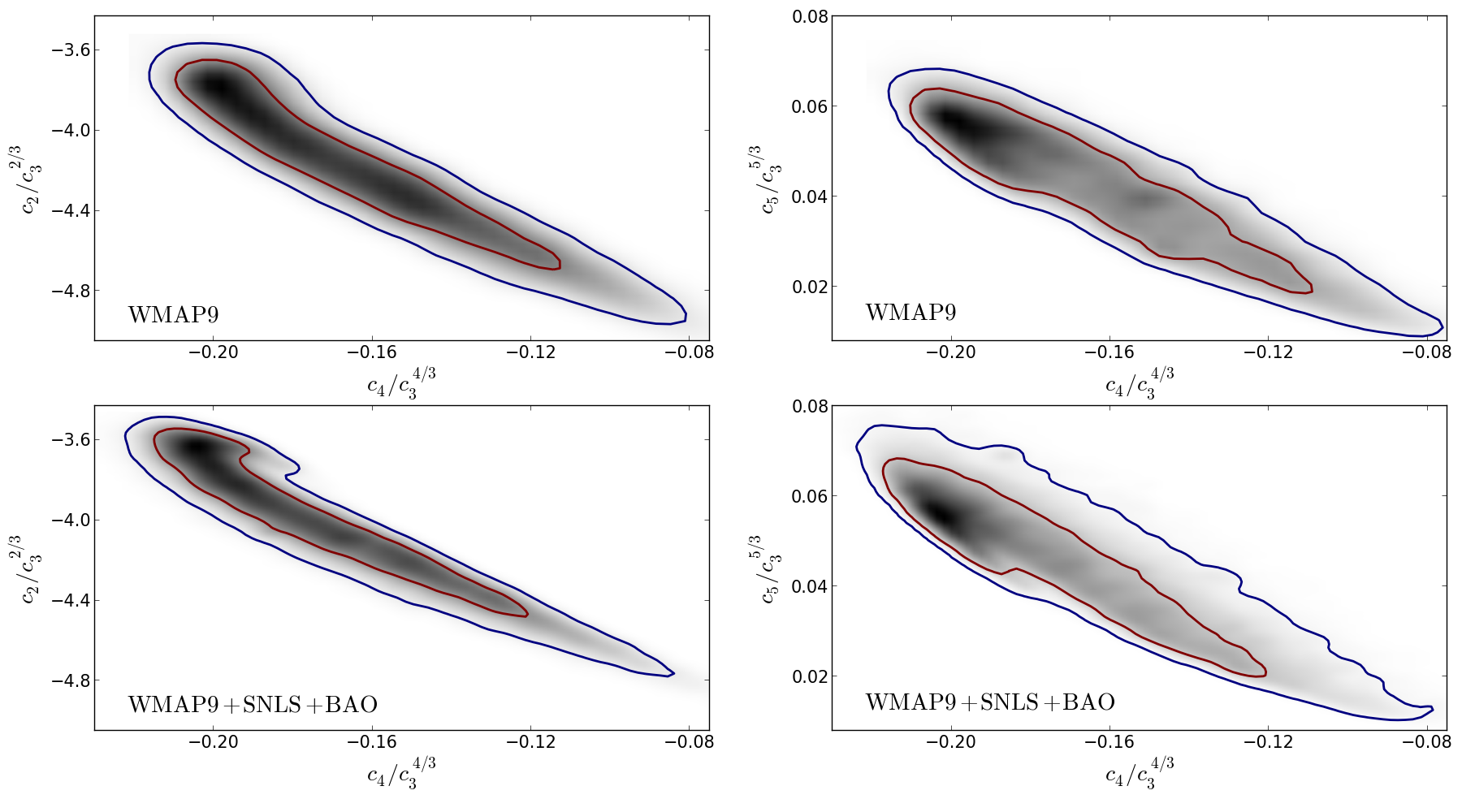}
	\caption{(Color online) Marginalized two-dimensional posterior distributions and respective 68$\%$ and 95$\%$ contour limits obtained for the Galileon sector of the parameter space with WMAP9 data alone (top panels) and the combined WMAP9+SNLS+BAO datasets (bottom panels). The shading represents the distribution, where darker regions mean higher probability density. The posterior distribution and the respective contours were smoothed using a Gaussian filter that did not change the underlying results. In these chains the parameter $c_3$ was held fixed at $c_3 = 10$ and $\dot{\bar{\varphi}}_i$ was allowed to take only positive values.}
\label{galileon-constraints}
\end{figure*}

\begin{figure*}
	\centering
	\includegraphics[scale=0.39]{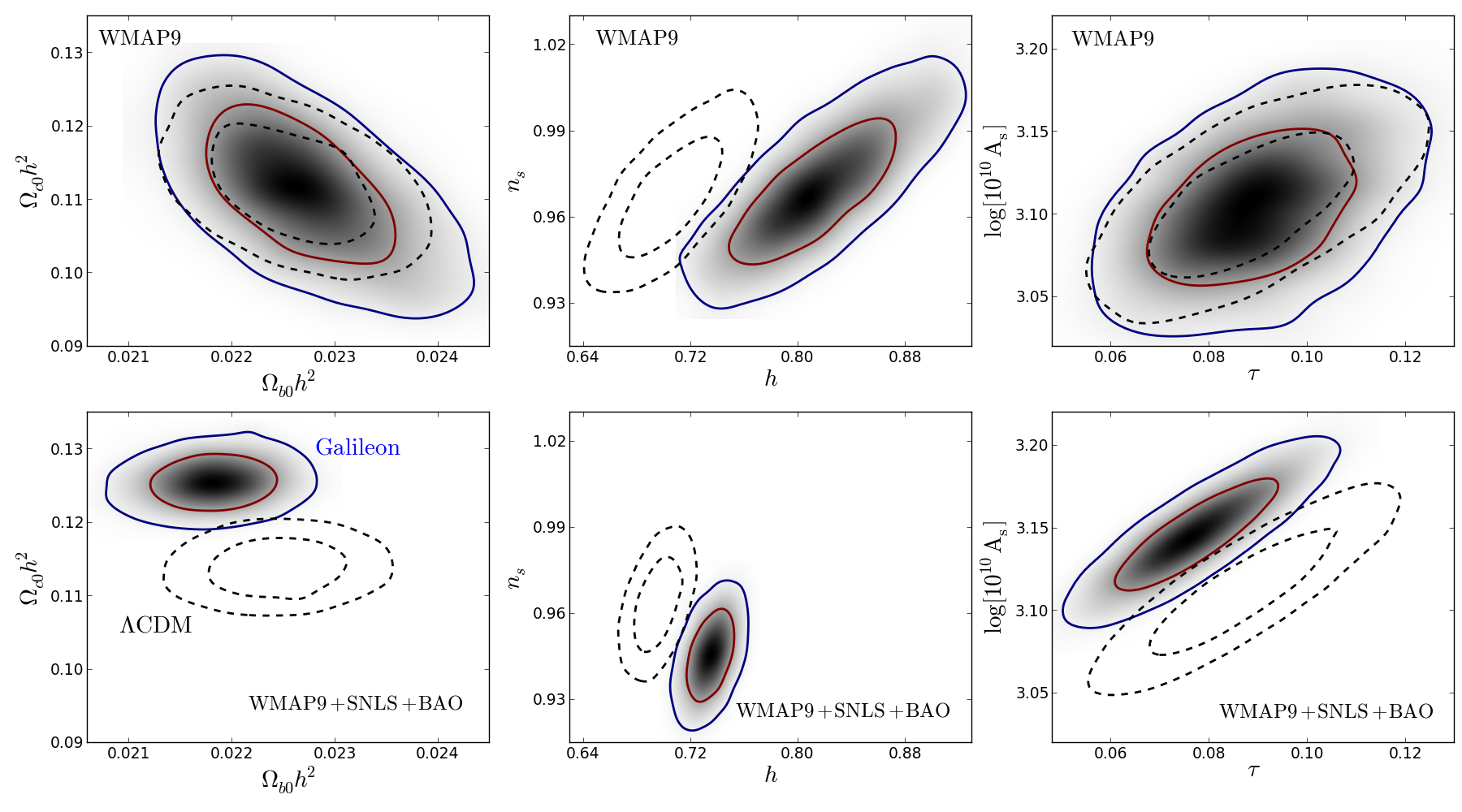}
	\caption{(Color online) Same as Fig.~\ref{galileon-constraints} but for the cosmological sector of the parameter space. The corresponding contours obtained for the $\Lambda$CDM model are also shown for comparison (dashed contours). The scalar amplitude at recombination $A_s$ refers to a pivot scale $k = 0.02 \rm{Mpc}^{-1}$.}
\label{cosmological-constraints}
\end{figure*}

\begin{figure*}
	\centering
	\includegraphics[scale=0.38]{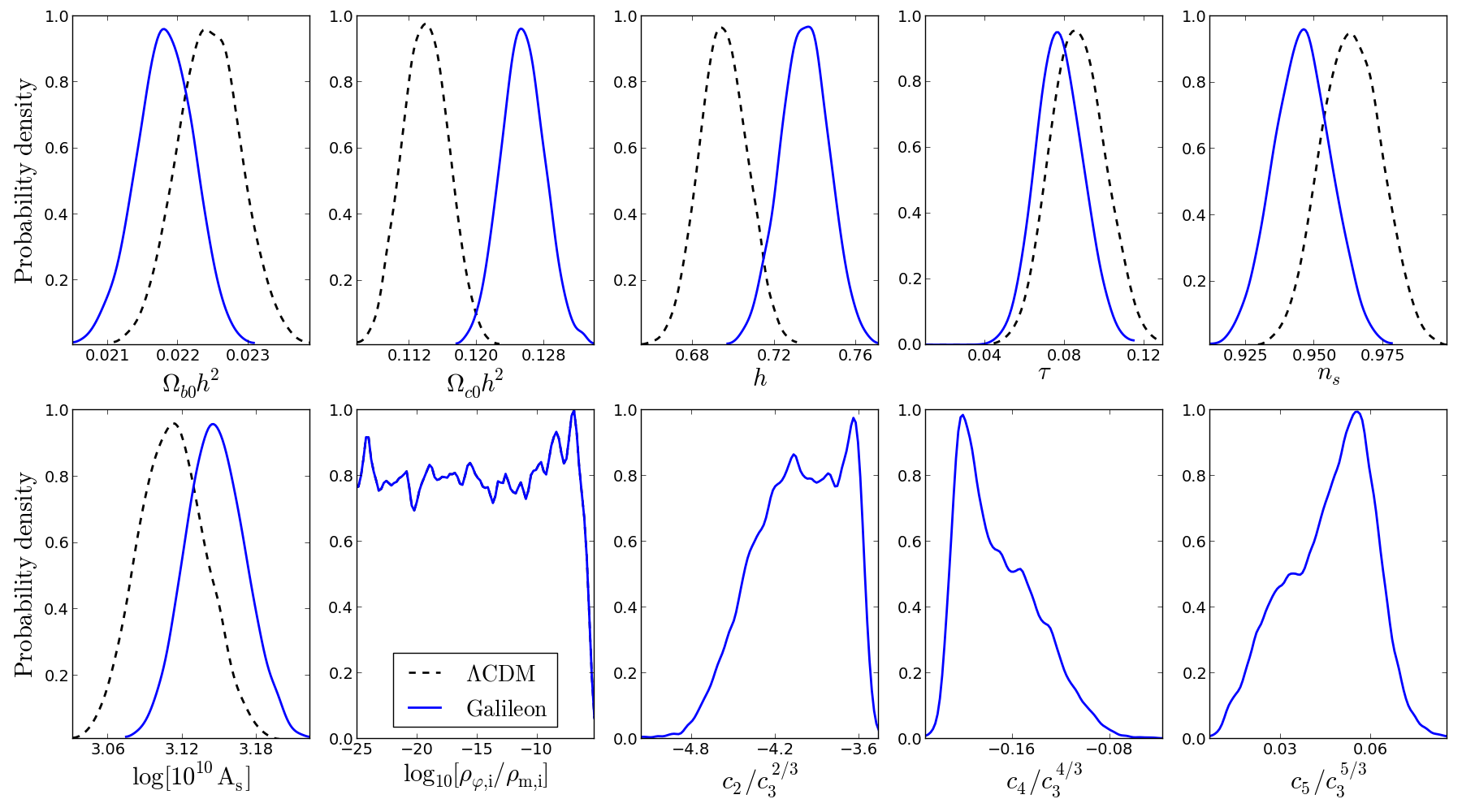}
	\caption{(Color online) Marginalized one-dimensional distributions obtained for the combined WMAP9+SNLS+BAO dataset (solid blue).  The distributions obtained for the $\Lambda$CDM model are also shown for comparison (dashed black curves). The scalar amplitude at recombination $A_s$ refers to a pivot scale $k = 0.02 \rm{Mpc}^{-1}$. In these chains the parameter $c_3$ was held fixed at $c_3 = 10$ and $\dot{\bar{\varphi}}_i$ was allowed to take only positive values.}
\label{panel-1D}
\end{figure*}
In Fig.~\ref{galileon-constraints} we show the marginalized two-dimensional likelihood distributions and the corresponding $68\%$ and $95\%$ contour limits of the Galileon sector of the parameter space using WMAP9 data (top panels) and the combined WMAP9+SNLS+BAO data (bottom panels). Figure~\ref{cosmological-constraints} shows the same but for the cosmological sector of the parameter space and the dashed black lines correspond to the $68\%$ and $95\%$ confidence limits for constraints on the $\Lambda$CDM model. We also show in Fig.~\ref{panel-1D} the marginalized one-dimensional distributions for constraints using the combined WMAP9+SNLS+BAO data. We point out that the constraints we derive for the $\Lambda$CDM model are in good agreement with those presented in the WMAP 9yr final result paper \cite{Hinshaw:2012fq} (c.f. Tables \ref{table-mean} and \ref{table-max}).

As we discussed above, fixing one of the Galileon parameters breaks the scaling degeneracy of the Galileon sector of the parameter space, and therefore, we see in Fig.~\ref{galileon-constraints}, that sensible constraints can now be derived. In particular, the parameters $c_2/c_3^{2/3}$, $c_4/c_3^{4/3}$ and $c_5/c_3^{5/3}$ are constrained to lie in a small region of the parameter space, very close to the theoretically unstable portion of the parameter space (see Fig.~\ref{figure-tensions} below)\footnote{In \cite{Appleby:2012ba, Neveu:2013mfa}, the authors have also found that the high-likelihood regions tend to lie close to the regions associated with ghosts or Laplace instabilities.}. The shape of the contours shows also that these parameters are correlated to some extent. In Fig.~\ref{panel-1D}, we see that the distribution of $\bar{\rho}_{\varphi, i} / \bar{\rho}_{m,i}$ is essentially flat, up to some statistical noise, but it decays very sharply for $\bar{\rho}_{\varphi, i} / \bar{\rho}_{m,i} \gtrsim -6$. This indicates that the upper bound of this parameter can play an important role in the physics of the Galileon \cite{Barreira:2012kk}, as we discuss below.

The constraints on the Galileon sector of the parameter space do not change substantially when adding the SNLS and BAO data to the WMAP9 data. However, the same does not happen in the cosmological sector, in which case the preferred values differ quite significantly between the two data combinations, as shown in Fig.~\ref{cosmological-constraints}. In particular, when adding the SNIa and BAO constraints, the Galileon model prefers higher values of the total matter density $\Omega_m{h}^2 \sim 0.145$, as opposed to $\Omega_m{h}^2 \sim 0.135$ for WMAP9 alone (see Tables \ref{table-mean} and \ref{table-max}). The amplitude of the primordial fluctuations, $A_s$, also increases going from WMAP9 only to the combined dataset, but the preferred values of the expansion rate $h$, of the spectral index $n_s$ and of the optical depth to reionization, $\tau$, become smaller. 

It is also interesting to compare how the constraints of the Galileon and $\Lambda$CDM models respond to different datasets. Here, we note that when using only the WMAP9 data, the Galileon and the $\Lambda$CDM models prefer more or less the same regions of parameter space with the exception of the Hubble rate ${h}$ which is larger in the Galileon model, by just over $1\sigma$. However, adding the SNIa and the BAO data leads to deviations in all of the cosmological parameters whose marginalized two-dimensional distributions can differ by more than $2\sigma$. More specifically, when using the combined dataset, one sees in Fig.~\ref{cosmological-constraints} that the Galileon model prefers more matter in the Universe, despite preferring slightly less baryons, and higher values of  the parameters $h$ and $A_s$ relative to $\Lambda$CDM. The parameters $n_s$ and $\tau$ tend to be smaller in the Galileon model than in $\Lambda$CDM. The discrepancy between the constraints of the Galileon and $\Lambda$CDM models for these cosmological parameters highlights the importance of allowing all of the cosmological parameters to vary when extracting numerical constraints, instead of fixing them to their best fit values in $\Lambda$CDM. For instance, note that the best-fitting values of $h$ to the combined WMAP9+SNLS+BAO dataset are closer in the Galileon model than in $\Lambda$CDM to the independent determinations of the present-day expansion rate $h = 0.742 \pm 0.036$ \cite{Riess:2009pu}, $h = 0.744 \pm 0.025$ \cite{Riess:2011yx} and $h = 0.743 \pm 2.1$ \cite{Freedman:2012ny}.

The Galileon model is able to fit the full WMAP9 data better than $\Lambda$CDM with a difference in $\chi^2$ of $\Delta\chi^2_{\rm{WMAP9}} \approx -1.8$ (see Fig.\ref{cmb-bf}; negative values of $\Delta\chi^2$ indicate the Galileon is favored). On the other hand, once the SN and the BAO data are added, then the Galileon becomes less favored with a difference of $\Delta\chi^2_{\rm{WMAP9+SNLS+BAO}} \approx 8.6$. This unveils a tension between the CMB data and the geometrical tests in the Galileon model and it is worth trying to understand the reasons behind it.

It is non-trivial that the constraints on the Galileon parameters barely change between the two data combinations. In \cite{Barreira:2012kk, Nesseris:2010pc, Appleby:2011aa, Appleby:2012ba}, it was shown that the observations that only probe the background evolution of the Universe typically prefer larger values of the initial density of the Galileon field because these minimize the strength and duration of the phantom evolution in the near past, therefore causing $w$ to approach the de Sitter attractor $w = -1$ sooner. However, Fig.\ref{galileon-constraints} shows that adding the SNLS and the BAO data to the WMAP9 data does not lead to a significant increase in the upper bound of $\bar{\rho}_{\varphi, i} / \bar{\rho}_{m,i}$ (see Table \ref{table-mean}). This is because the increase of the initial density has a larger impact on the time evolution of the gravitational potential than on the evolution of the background expansion rate. In \cite{Barreira:2012kk} we showed that if the initial density of the field is too large, $\bar{\rho}_{\varphi, i} / \bar{\rho}_{m,i} \gtrsim 10^{-5}$ at $z_i = 10^6$, then this generally leads to strong time variations of the gravitational potential and hence to an unacceptably large ISW effect. This is corroborated by the very sharp decrease in the posterior distribution towards large values of $\bar{\rho}_{\varphi, i} / \bar{\rho}_{m,i}$, as seen in Figs.~\ref{galileon-constraints} and \ref{panel-1D}. As a result, the ISW effect constrains the upper bound of this parameter so well that the constraint is barely changed when the background data is also considered. As we will see below, the parameters $c_4$ and $c_5$ are also very tightly constrained by the CMB data (c.f. Fig.\ref{figure-tensions}).

In \cite{Barreira:2012kk}, we have also shown that, if $\bar{\rho}_{\varphi, i} / \bar{\rho}_{m,i}$ is sufficiently small, then different values will have no visible impact on the CMB power spectrum.  The high-redshift evolution of $w$ will still be sensitive to $\bar{\rho}_{\varphi, i} / \bar{\rho}_{m,i} \lesssim 10^{-5}$, but this dependency does not propagate to the expansion rate, which is only sensitive to $w$ when the dark energy becomes the dominant component at low redshift. This is what makes the distribution of $\bar{\rho}_{\varphi, i} / \bar{\rho}_{m,i}$, effectively, unbounded from below\footnote{Keeping in mind the positive sign of the energy density. Sufficiently negative values of the energy density are typically associated with ghosts.}, as one can see in Fig.\ref{panel-1D}.

Since the Galileon parameters are very well constrained by the ISW effect, the main consequence of adding the SNIa and BAO data to the CMB data would be to "tune" the total matter density in the Universe (which increases) and the present-day expansion rate (which decreases). This modifies the time evolution of $H(a)$, but in a way that preserves the positions of the acoustic peaks of the CMB. This can be seen by comparing the values for the age of the Universe in Tables \ref{table-mean} and \ref{table-max}, which barely change between the two data combinations. However, these changes will have an impact on the amplitude of the CMB power spectrum, which causes the remaining cosmological parameters to explore the possible degeneracies in order to keep as good a fit. As a result, they deviate from their best-fitting values obtained using the WMAP9 data alone. This "compensation" is, however,  not perfect and the model is forced to reach a compromise between the regions of parameters favored by WMAP9 and SNLS+BAO. Overall, the net result is a poorer fit to the combined dataset. This tension between the background constraints and the ISW effect is similar to that highlighted in Refs.~\cite{Appleby:2011aa, Neveu:2013mfa}. However, here we want to emphasize that this is a tension between the cosmic background evolution and the {\it cumulative effect of the time variation of the lensing potential}, instead of one between the background evolution and the {\it summed effect of the Newtonian potential} \footnote{In the conformal Newtonian gauge with the line element given by ${\rm d}s^2=a^2\left[(1+2\Psi){\rm d}\tau^2-(1-2\Phi){\rm d}{\mathbf x}^2\right]$, the ISW effect probes $(\dot{\Psi}+\dot{\Phi})/2=\dot{\phi}$, where $\phi$ is the lensing (Weyl) potential, while particle dynamics, and therefore the growth rate, is sensitive to $\Psi$. Because of the nonzero anisotropic stress, Eq.~(\ref{perturbed4}), these two potentials can be different in the Galileon model.}. In addition, this tension shows up at the largest length scales relevant for the ISW effect, where we are confident that the Vainsthein screening plays a negligible role so that there is no concern about the validity of linear perturbation theory.

\subsection{Cosmology of the best fit}\label{best fit}

\begin{figure}
	\centering
	\includegraphics[scale=0.385]{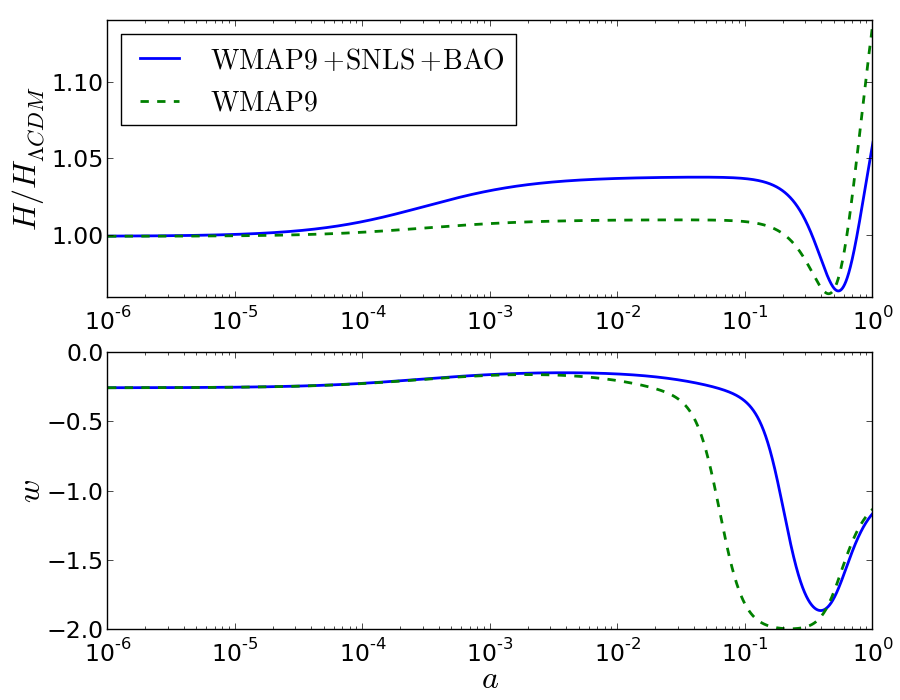}
	\caption{(Color online) Time evolution of the ratio of the Hubble expansion rates of the Galileon and $\Lambda$CDM models, $H/H_{\Lambda\rm{CDM}}$ (top panel) and of the Galileon field equation of state parameter $w$ (bottom panel) for the maximum likelihood  points of the chains using the WMAP9 data alone (dashed green) and the combined dataset WMAP9+SNLS+BAO (solid blue). The $\Lambda$CDM model used in the ratio of the upper panel is the best-fitting model obtained using the WMAP9+SNLS+BAO combined dataset.}
\label{H-w-bf}\end{figure}

\begin{figure*}
	\centering
	\includegraphics[scale=0.545]{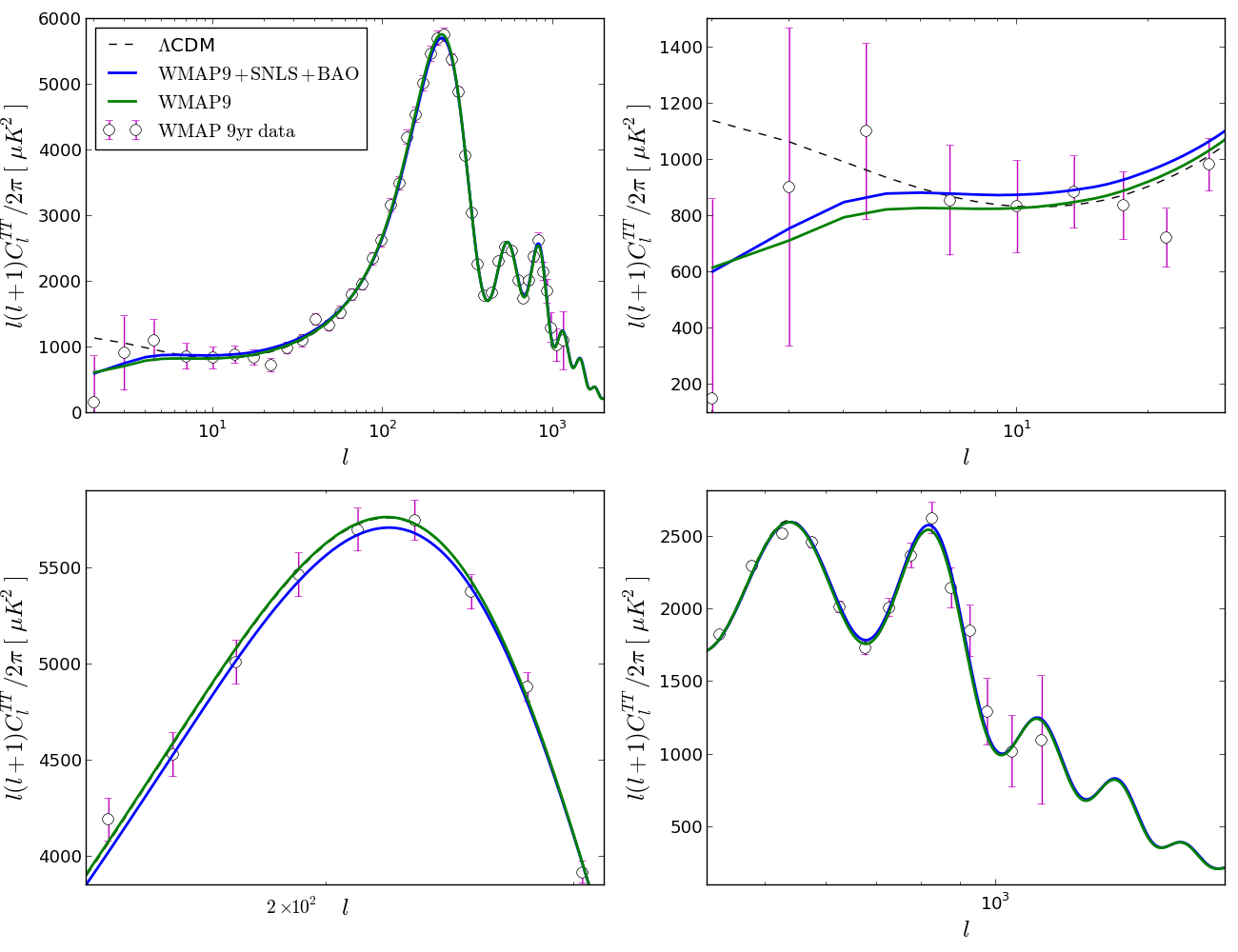}
	\caption{(Color online) CMB angular power spectra for the maximum likelihood points of the chains using the WMAP9 data alone (solid green) and the combined WMAP9+SNLS+BAO dataset (solid blue). The power spectrum of the best-fitting $\Lambda$CDM model to the WMAP9 data alone is also shown for comparison (dashed black). The WMAP 9yr data is represented by the open circles with errorbars \cite{Hinshaw:2012fq}. The top right, bottom left and bottom right panels show zooms into the regions most relevant for the ISW effect, the first acoustic peak and the higher CMB peaks, respectively, of the top left panel. In the top left and top right panels at $l = 4$, from top to bottom, the curves correspond to $\Lambda$CDM, the best-fitting Galileon model using the combined WMAP+SNLS+BAO dataset and the best-fitting Galileon model using the WMAP9 data alone, respectively. In the bottom left panel, the $\Lambda$CDM curve is indistinguishable from the curve of the best-fitting Galileon model to the WMAP9 data alone and at the CMB peak position, the curve of the best-fitting Galileon model to the combined WMAP9+SNLS+BAO dataset is the one with lower amplitude. In the bottom right panel, the $\Lambda$CDM curve is indistinguishable from the curve of the best-fitting Galileon model to the WMAP9 data alone, but at the third CMB peak position (the second one shown in the panel), the curve with higher power is the curve of the best-fitting Galileon model to the combined dataset. In the rest of the bottom right panel the three curves are indistinguishable.}
\label{cmb-bf}\end{figure*}

\subsubsection{Background evolution}

In Fig.~\ref{H-w-bf} we show the time evolution of the ratio $H/H_{\Lambda \rm{CDM}}$ and $w$ of the best-fitting points (shown in Table \ref{table-max}) obtained for the constraints with WMAP9 data alone (dashed green) and the combined WMAP9+SNLS+BAO data (solid blue). Here, the $\Lambda$CDM model is the best-fitting model to the combined WMAP9+SNLS+BAO data. The expansion rate of the two best-fitting Galileon models agrees with $\Lambda$CDM during the radiation dominated era since these models all have the same energy density of radiation. However, in the matter era, the best-fitting Galileon model to the combined dataset has an expansion rate which can be up to $\approx 4\%$ higher than both the best fit Galileon model of the WMAP9 data and the $\Lambda$CDM model, as consequence of the also higher total matter density (see Fig.~\ref{cosmological-constraints} and Tables \ref{table-mean} and \ref{table-max}). At the end of the matter era, the expansion rates of the two Galileon models first decrease to be approximately $4\%$ smaller than $\Lambda$CDM, and then start increasing towards the present-day, at $a \approx 0.5$. This late growth is more pronounced in the former since the value of the present-day expansion rate, $h$, is larger.

In the best-fitting Galileon model to the combined WMAP9+SNLS+BAO dataset, the phantom evolution $w < -1$ starts closer to the present-day and is also less pronounced than in the best-fitting Galileon model to the WMAP9 data alone. This is mainly due to the larger value of $\rm{log}_{10}\left[\rho_{\varphi, i} / \rho_{m,i}\right]$ in the former relative to the latter (c.f. Table \ref{table-max}). As we showed in Fig.~1 of \cite{Barreira:2012kk} (see also \cite{Appleby:2011aa}), the lower $\rho_{\varphi, i} / \rho_{m,i}$ the more negative the values of $w$ will be. This is because lower values of the Galileon density in the past will force the energy density to grow more drastically ($w<-1$) closer to today when the field starts to be driven towards the de Sitter attractor evolution. Recall, however, that the low redshift background data has little power to increase the initial Galileon density, since this parameter is very tightly constrained by the ISW effect (c.f. Fig.~\ref{galileon-constraints}).

It is interesting to compare the typical time evolution of the Galileon equation of state parameter with those reconstructed from the observational data in \cite{Zhao:2012aw} (see also \cite{Crittenden:2005wj, Crittenden:2011aa}). By using non-parametric Bayesian statistical techniques these authors have found that the currently available data favors dynamical dark energy models. In particular, the preferred evolution of $w$ is such that it crosses the `phantom' line at $z \sim 0.5$ from $w > -1$, and approaches $w = -1$ from below at around the present time, which is not too different from what happens in the evolution of the background in the best-fitting Galileon model. The interested reader might wish to compare Fig.~\ref{H-w-bf} of this paper with Fig.~1 of \cite{Zhao:2012aw}.

\subsubsection{Cosmic Microwave Background}

Fig.~\ref{cmb-bf} shows the CMB temperature angular power spectrum of the best-fitting distribution values of the chains constrained using the WMAP9 data alone (solid green) and the combined WMAP9+SNLS+BAO data (solid blue)  (c.f.~Table \ref{table-max}). The top right, bottom left and bottom right panels zoom into the three main regions of interests in the top left panel, which are, respectively, the low-$l$ region which is most relevant to the ISW effect, the region of the first acoustic peak and the region of the higher peaks. For comparison, we also show the prediction of the best-fitting $\Lambda$CDM model constrained by the WMAP9 data alone, which we show as the open circles with errorbars. We can see that the Galileon model gives a very good fit to the measured angular power of the CMB perturbations. For $l \gtrsim 100$, the best-fitting Galileon models give essentially the same prediction as $\Lambda$CDM, although the best-fitting Galileon model to the combined dataset (solid blue) predicts slightly more power in the interval of $l$ that lies between the second and the third acoustic peaks.

The most pronounced feature in the CMB predictions of the best-fitting Galileon models is, however, seen in the low-$l$ region of the spectrum where the ISW effect dominates. In \cite{Barreira:2012kk}, we pointed out that in Galileon models the gravitational (Weyl) potential can undergo dramatic time variations, leading to a significant increase in the amplitude of the low-$l$ CMB power. However, the results here show that the opposite is also possible, as it is the case of the best-fitting Galileon models. Indeed, even though the time variation of the gravitational potential can be quite strong, it can also cancel out when integrated, yielding a weaker ISW effect for the best-fitting Galileon models, which even exhibit an inflection in the shape of the CMB power spectrum at $l \approx 6$. This is in clear contrast with the $\Lambda$CDM curve, which keeps growing towards lower $l$, and such an interesting result can in principle be used to distinguish between the Galileon model and other dark energy or modified gravity models. It is well known that the CMB signals at such low $l$ are severely affected by cosmic variance (note the large error bars on the WMAP data points), but nevertheless the CMB data does seem to prefer the Galileon models which exhibit such an inflection in the ISW power over those which do not. Note that this inflection is not a generic feature of the Galileon parameter space whose rich phenomenology allows it to have CMB power spectra which are also very similar to $\Lambda$CDM, as shown in \cite{Barreira:2012kk}.

To see more clearly the effect of varying the $c_n$ parameters, in the top right panel of Fig.~\ref{figure-tensions} we plot the CMB power spectrum for the parameter points highlighted in the top left panel. In the latter, we zoom into a region of the $c_4 - c_5$ plane constrained with the WMAP9+SNLS+BAO data (bottom right panel of Fig.\ref{galileon-constraints}), to show the accepted points by the MCMC algorithm rather than the contours of the posterior distribution. This enables us to resolve the fine details of the constrained parameter space better than by evaluating the likelihood contour, which always requires some interpolation and smoothing. This is particularly useful in the Galileon model, whose parameter space can have a quite detailed shape. Indeed, we can see a gap between the accepted (cyan) and unstable (grey) points, filled with low-likelihood points for which the ISW effect is too strong. Furthermore, there is a `hole' of low-likelihood points near the boundary of accepted points (where the point $B_1$ is located), which is not visible in the contour representation. Points $A_1$, $A_2$, $A_3$, $B$ and $B_1$ are obtained from the best-fitting point $A$ by keeping all the parameters fixed except for $c_4$ and $c_5$, which are varied to explore the different parts of the $c_4-c_5$ plane: $A_1$ and $A_2$ lie in the low-likelihood gap between the main sampled region and the unstable portion; $A_3$ is  a typical accepted point; $B$ is another accepted point which gives a fit nearly as good as $A$; and $B_1$ is a rejected point in the low-likelihood `hole' mentioned above.

From Fig.~\ref{figure-tensions} we can see that going from the best-fitting point $A$ (green curve) inwards to the contour into another point still deep inside the 68$\%$ confidence limit (point $A_3$, red line), we end up with nearly the same CMB power spectrum. On the other hand, moving outwards from the contour into points $A_1$ (black line) and $A_2$ (blue line) we get a sudden increase in the ISW power which is enough to render these models incompatible with the data and, therefore, gives this region a low likelihood (in fact, not a single point was sampled by the MCMC search in the neighborhood of the points $A_1$ and $A_2$). One sees that the ISW effect plays a key role in constraining the Galileon model, as already anticipated in our earlier work \cite{Barreira:2012kk}: the allowed region in the $c_4-c_5$ plane corresponds to the values of these parameters for which the time evolution of the gravitational potential is mild enough that it yields small power on the largest angular scales (keeping in mind also that $\bar{\rho}_{\varphi, i}$ cannot be too large). The expansion rate evolution of the points in this region is essentially the same. The very strong sensitivity to the Galilean parameters can also be seen by looking at the CMB power spectra of points $B$ (cyan line) and $B_1$ (magenta line). Point $B$ lies in a narrow layer of accepted points and gives an almost indistinguishable CMB spectrum compared to point $A$. Nevertheless, taking a small step towards the less favored region to point $B_1$ leads to a significant increase in the power for $l \lesssim 11$. Such an increase is however not as pronounced as that seen when going from point $A$ to points $A_1$ and $A_2$, which is why some points in that region were still sampled during the MCMC search.

The differences in the amplitude of the CMB spectra within the parameter space of the Galileon model show the advantage of using the full CMB data when constraining the Galileon model, over using only the WMAP distance priors as done in \cite{Nesseris:2010pc, Appleby:2012ba, Neveu:2013mfa}. For instance, using only the information encoded in the positions of the acoustic peaks of the CMB would result in nearly the same fit for all the models shown in Figs.~\ref{cmb-bf} and \ref{figure-tensions}, although, in fact, these can be very easily distinguished and, in some cases, completely ruled out by the full CMB data.

\subsection{Future constraints}

\begin{figure*}
	\centering
	\includegraphics[scale=0.545]{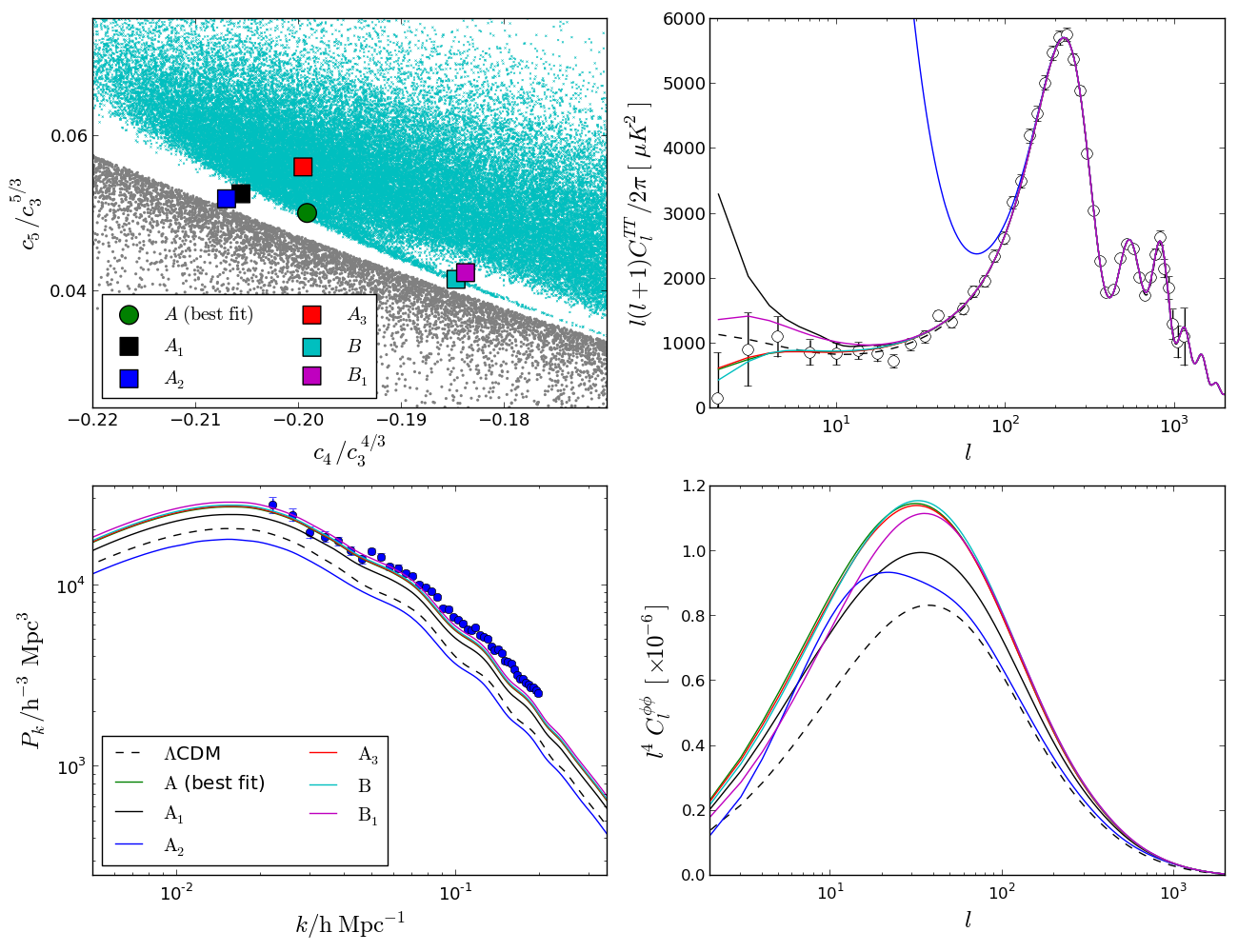}
	\caption{(Color online) The top left panel shows the points accepted by the MCMC algorithm (cyan dots) after the burn-in period for a zoomed region of the $c_4-c_5$ plane constrained with the combined WMAP9+SNLS+BAO data. The highlighted points have the same parameters as the best-fitting Galileon model (point $A$, green circle) except for $c_4$ and $c_5$ which were chosen to explore other regions of the $c_4 - c_5$ plane. The grey dots show the points sampled by the MCMC algorithm but that were rejected by being either associated with ghost or Laplace instabilities. The figure also shows the CMB power spectra (top right panel), linear matter power spectra at redshift $\bar{z}_{\rm LRG} = 0.31$ (bottom left panel) and angular power spectrum of the weak lensing potential (bottom right panel) for the Galileon models corresponding to the highlighted points in the top left panel and for the best-fitting $\Lambda$CDM model (dashed line) of the combined WMAP9+SNLS+BAO dataset. We also plot the WMAP 9yr data (open circles) \cite{Hinshaw:2012fq} and the SDSS-DR7 LRG host halo power spectrum data (blue dots) \cite{Percival:2009xn} in the top right and bottom left panels, respectively. $\bar{z}_{\rm LRG}$ is the mean redshift of the LRG sample. In the top right panel, at $l = 2$, for top to bottom the lines correspond, respectively, to the points $A_2$ (goes over the plot limit), $A_1$, $B_1$, $\Lambda$CDM, $A_3$, $A$ and $B$. In the bottom left plot, from bottom to top, the lines correspond to point $A_2$, $\Lambda$CDM, point $A_1$ and are indistinguishable from there on. In the bottom left panel, from top to bottom, at the peak position of the $\Lambda$CDM curve (dashed black) the lines correspond to the points $B$, $A$, $A_3$, $B_1$, $A_1$, $A_2$ and $\Lambda$CDM, respectively.}
\label{figure-tensions}
\end{figure*}

The bottom left and bottom right panels of  Fig.~\ref{figure-tensions} show the power spectra for the linear matter density fluctuations and the weak lensing potential, respectively, corresponding to the models highlighted in the top left panel. In the bottom left we also show a recent estimate of the power spectrum of luminous red galaxies (LRGs) of the DR7 from SDSS \cite{Percival:2009xn}. The predictions of the Galileon power spectrum are shown at $\bar{z}_{\rm LRG} = 0.31$, which is the median redshift of the LRGs. This way one does not need to account for the growth factor to compare the theoretical spectrum with the observed one. However, there are at least three other effects that need to be considered, including baryonic bias effects, redshift space distortions and nonlinearities. The latter includes the familiar mode coupling that develops between perturbations on different scales, but also, and this is particularly important in modified gravity theories, possible screening effects (the Vainshtein screening for the case of Galileon models) which can render linear perturbation theory invalid on scales where it is usually taken to be a good approximation \cite{Brax:2010gi, Brax:2011ja, Brax:2012nk, Brax:2012nk, Li:2012by}. We stress once again that this is the ultimate reason why we chose to leave growth rate and clustering measurements out of the constraints in this paper, and conclude that a better understanding of the screening in Galileon models is necessary for reliable constraints using these datasets.

However, in case the Vainshtein screening mechanism does not affect the linear predictions too much, a brief study of the linear matter power spectrum can still shed light on our understanding of the model and indicate what we can expect when combining the different datasets. For this, one still needs to consider galaxy bias and redshift space distortion effects which link the theoretical matter power spectrum to the observed galaxy power spectrum. Galaxy bias arises because the baryons, compared with the dark matter particles, experience additional non-gravitational interactions which alter the spatial distribution of galaxies relative to dark matter. Redshift space distortions result from the peculiar motions of galaxies, which add to the Hubble flow, and result in systematic changes in the clustering power and pattern of the observed galaxies \cite{Jennings:2011}. 

In the simplest cases, these effects are modeled by a scale-independent change in the amplitude of the spectra, though numerical simulations show that in reality these effects are scale-dependent, especially for highly clustered objects \cite{Angulo:2008, Jennings:2011}. There is evidence which suggests that LRGs are biased tracers of the underlying dark matter distribution \cite{Wake:2008, Zheng:2009, Sawangwit:2011, Kulkarni:2007, Marin:2011, Percival:2009xn}, which means that the observed galaxy spectrum should have a higher amplitude than the dark matter spectrum predicted by linear theory, independently of the theory of gravity. This simple requirement already puts many of the Galileon models (e.g, most of those shown in Fig.~\ref{figure-tensions}) at odds with the observations. The only exceptions are the points $A_1$ and $A_2$ whose linear matter power spectra are the closest to the $\Lambda$CDM prediction. In particular, point $A_2$ predicts a lower clustering power than the standard $\Lambda$CDM model. However, these points are disfavored by the CMB data because they produce too much ISW power (top right panel of Fig.~\ref{figure-tensions}). This unveils yet another tension in the Galileon model, in addition to that between the background evolution and the ISW effect and the one pointed out in \cite{Appleby:2012ba}. In particular, one sees in Table \ref{table-mean} that the mean value of the variance of the power spectrum at $k = 1/8 h \rm{Mpc}^{-1}$, $\sigma_8$, is $\sigma_8 \approx 1$ for the Galileon model, as opposed to $\sigma_8 \approx 0.8$ for $\Lambda$CDM. With these results we can anticipate that, providing linear theory holds on the scales of the observations, complementing our study with clustering measurements will result in even tighter constraints than those quoted in Tables~\ref{table-mean} and \ref{table-max}. The only salvation would arise if any modifications introduced by the Vainshtein mechanism are able to bring the clustering predictions closer to the observations.

In \cite{Barreira:2012kk} we saw that the power spectrum of the weak lensing (Weyl) potential $\phi$ can also lead to very distinctive observational signatures of the Galileon model. The weak lensing signal is the integrated effect of the Weyl gravitational potential along the line-of-sight from today to the epoch of recombination \cite{Lewis:2006fu, Zhao:2008bn}. In Fig.~\ref{figure-tensions} we see that the angular power of the lensing potential in the Galileon model is higher than $\Lambda$CDM on all scales for all of the models shown, which includes models whose CMB power spectrum fit the WMAP9 data sufficiently well. This once again provides another potentially powerful way of discriminating between the Galileon and the standard $\Lambda$CDM model. Moreover, in the case of the weak lensing power spectrum, just as in the CMB case, the length scales are large enough for one to not worry about whether or not the Vainshtein effect has an impact on the validity of linear perturbation theory. However, the currently available data still has little constraining power, even for the $\Lambda$CDM model, and therefore no decisive conclusions can be drawn for now. Nevertheless, this shows that weak lensing measurements, if sufficiently accurate, may provide a promising way of constraining the Galileon model, which can be in many aspects representative of other variations of the general Horndeski model. For this, it would be interesting to forecast how accurate future lensing measurements would need to be, like the ones of Euclid and Planck, in order to distinguish the Galileon model from $\Lambda$CDM.

\begin{table*}
\caption{Mean values and uncertainties ($1\sigma$ level) of the marginalized one-dimensional distributions of the full cosmological parameter space for the Galileon (except $\rm{log}_{10}\left[\rho_{\varphi, i} / \rho_{m,i}\right]$\footnote{Since the distribution of $\rm{log}_{10}\left[\rho_{\varphi, i} / \rho_{m,i}\right]$ is unbounded from below, it does not make sense to quote its mean. Instead, we quote the highest value sampled by the chains to give an estimate where the distribution cuts-off.}) and $\Lambda$CDM models using WMAP9 alone and the combined WMAP9+SNLS+BAO data. The scalar amplitude at recombination $A_s$ refers to a pivot scale $k = 0.02 \rm{Mpc}^{-1}$. The subscript "$_i$" refers to quantities evaluated at $z = z_i = 10^6$. We also show the mean values of the age of the Universe and of $\sigma_8$ at redshift $z = 0$, which are derived parameters.}
\begin{tabular}{@{}lccccccccccc}
\hline\hline
\\
Parameter  & \ \ Galileon & \ \ Galileon  & \ \ $\Lambda$CDM & \ \ $\Lambda$CDM & \\
                       & \ \ (WMAP9) & \ \ (WMAP9+SNLS+BAO) & \ \ (WMAP9) & \ \  (WMAP9+SNLS+BAO) & \\
\\
\hline
\\
$\Omega_{b0}{h}^2$  							&\ \ $0.0227_{-0.0006}^{+0.0007}$ & \ \ $0.02181 \pm 0.00041$ & \ \ $0.02259_{-0.00056}^{+0.00054}$ & \ \ $0.02244 \pm 0.00045$ & \\
\\
$\Omega_{c0}{h}^2$  							&\ \ $0.111_{-0.008}^{+0.007}$ & \ \ $0.1255 \pm 0.0026$ & \ \ $0.112_{-0.055}^{+0.053}$ & \ \ $0.1139 \pm 0.0028$ & \\
\\
${h}$  									&\ \ $0.817 \pm 0.04$ & \ \ $0.735 \pm 0.012$ & \ \ $0.705 \pm 0.027$ & \ \ $0.695 \pm 0.012$ & \\
\\
$n_s$  									&\ \ $0.971 \pm 0.018$ & \ \ $0.945 \pm 0.011$ & \ \ $0.967 \pm 0.015$ & \ \ $0.963 \pm 0.011$ & \\
\\
$\tau$  									&\ \ $0.0898_{-0.0068}^{+0.0062}$ & \ \ $0.0775_{-0.0057}^{+0.0052}$ & \ \ $0.0889_{-0.0069}^{+0.0060}$ & \ \ $0.0867_{-0.0067}^{+0.0062}$ & \\
\\
$\rm{log}_{10}\left[ 10^{10}A_s \right]$  				&\ \ $3.104_{-0.033}^{+0.034}$ & \ \ $3.147_{-0.024}^{+0.025}$ & \ $\ 3.107_{-0.029}^{+0.031}$ & \ \ $3.112 \pm 0.028$ & \\
\\
$\rm{log}_{10}\left[\rho_{\varphi, i} / \rho_{m,i}\right]$  	&\ \ $-5.18^a$ & \ \ $-5.44^a$ & \ \ $---$ & \ \ $---$ & \\
\\
$c_2 / c_3^{2/3}$  							&\ \ $-4.21_{-0.38}^{+0.39}$ & \ \ $-4.04_{-0.34}^{+0.35}$ & \ \ $---$ & \ \ $---$ & \\
\\
$c_4/c_3^{4/3}$  								&\ \ $-0.161_{-0.36}^{+0.34}$ & \ \ $-0.171_{-0.032}^{+0.035}$ & \ \ $---$ & \ \ $---$ & \\
\\
$c_5/c_3^{5/3}$ 								 &\ \ $0.042_{-0.014}^{+0.016}$ & \ \ $0.046_{-0.017}^{+0.014}$ & \ \ $---$ & \ \ $---$ & \\
\\
\hline
\\
$\sigma_8\ (z = 0)$					  		&\ \ $0.938 \pm 0.038$ & \ \ $0.988 \pm 0.022$ & \ $0.810 \pm 0.026$ & \ \ $0.817 \pm 0.020$ & \\
\\
Age (Gyr)					  		&\ \ $13.49_{-0.18}^{+0.17}$ & \ \ $13.770_{-0.087}^{+0.086}$ & \ $13.75 \pm 0.13$ & \ \ $13.793 \pm 0.093$ & \\
\\
\hline
\hline
\end{tabular}
\label{table-mean}
\end{table*}

\begin{table*}
\caption{Maximum likelihood points in the MCMC chains together with the best-fitting value of $\chi^2 = -2\rm{log}P$ for the Galileon and $\Lambda$CDM models using WMAP9 alone and the combined WMAP9+SNLS+BAO data. The scalar amplitude at recombination $A_s$ refers to a pivot scale $k = 0.02 \rm{Mpc}^{-1}$. We also show the initial value of the Galileon field time derivative and of the age of the Universe. The subscript "$_i$" refers to quantities evaluated at $z = z_i = 10^6$.}
\begin{tabular}{@{}lccccccccccc}
\hline\hline
\\
Parameter  & \ \ Galileon & \ \ Galileon  & \ \ $\Lambda$CDM & \ \ $\Lambda$CDM & \\
                       & \ \ (WMAP9) & \ \ (WMAP9+SNLS+BAO) & \ \ (WMAP9) & \ \  (WMAP9+SNLS+BAO) & \\
\\
\hline
\\
$\chi^2$                                                               			         &\ \ $7556.87$ & \ \ $7989.97$ & \ \ $7558.64$ & \ \ $7981.42$ & \\
\\
$\Omega_{b0}{h}^2$                                				&\ \ $0.02233$ & \ \ $0.02178$ & \ \ $0.02238$ & \ \ $0.02237$ & \\
\\
$\Omega_{c0}{h}^2$                                 				&\ \   $0.116$ & \ \ $0.125$ & \ \ $0.113$ & \ \ $0.113$ & \\
\\
${h}$                                                                			&\ \   $0.789$ & \ \  $0.735$& \ \ $0.697$ & \ \ $0.693$ & \\
\\
$n_s$                                                                 			           &\ \    $0.957$ & \ \ $0.947$ & \ \ $0.966$ & \ \ $0.963$ & \\
\\
$\tau$                                                                     			&\ \   $0.0777$ & \ \ $0.0680$ & \ \  $0.0879$ & \ \ $0.0892$ & \\
\\
$\rm{log}_{10}\left[ 10^{10}A_s \right]$    			&\ \  $3.112$ & \ \ $3.127$ & \ \ $3.117$ & \ \ $3.116$ & \\
\\
$\rm{log}_{10}\left[\rho_{\varphi, i} / \rho_{m,i}\right]$         &\ \   $-9.15$ & \ \ $-6.51$ & \ \ $---$ & \ \ $---$ & \\
\\
$c_2 / c_3^{2/3}$                                                                                   &\ \    $-3.67$ & \ \ $-3.59$ & \ \ $---$ & \ \ $---$ & \\
\\
$c_4/c_3^{4/3}$                                                                                      &\ \ $-0.195$ & \ \ $0.199$ & \ \ $---$ & \ \ $---$ & \\
\\
$c_5/c_3^{5/3}$                                                                                     &\ \ $0.0485$ & \ \ $0.0501$ & \ \ $---$ & \ \ $---$ & \\
\\
\hline
\\
$\dot{\bar{\varphi}}_i c_3^{1/3}$					  		&\ \  $7.42\times10^{-15}$ & \ \ $2.31\times10^{-14}$ & \ $---$ & \ \ $---$ & \\
\\
Age (Gyr)					  		&\ \ $13.572$ & \ \ $13.778$ & \ $13.789$ & \ \ $13.819$ & \\
\\
\hline
\hline
\end{tabular}
\label{table-max}
\end{table*}

\section{Discussions and Conclusions}

\label{Conclusion}

We have studied and constrained, for the first time, the full parameter space of Galileon gravity models, which is composed of the six cosmological parameters $\Omega_{c0}, \Omega_{b0}, h, \tau, n_s, \log\left[10^{10}A_s\right]$; plus the Galileon parameters, using the latest observational data on SNIa, BAO and the full temperature anisotropy power spectrum of the CMB. This is the first time the full CMB data, rather than just the information encoded in the positions of the acoustic peaks, has been used to place constraints on the Galileon model. Our results were obtained using a modified version of the $\tt CAMB$ code, which solves for the cosmological background and linear perturbation evolutions \cite{Barreira:2012kk}. We implemented these solutions in the publicly available $\tt CosmoMC$ code to sample over the parameter space.  

We found that there is a set of scaling relations which can be used to change the Galileon parameters without changing the physics. We showed that when all of the parameters are allowed to vary, the scaling relations conspire to produce an unlimited degeneracy region in the Galileon parameter space along which the likelihood barely changes. A particularity of this degeneracy region is that it is symmetric in the signs of the parameters $c_3$, $c_5$ and $\dot{\bar{\varphi}}_i$. A similar scaling relation was also found in \cite{DeFelice:2010pv} (and more recently in \cite{Neveu:2013mfa}). However, in this paper we treat it in a way which allows us to directly constrain the physical parameters of the Lagrangian. Due to the scaling degeneracy, the dimensionality of the Galileon parameter space is reduced by one and we need only to constrain specific combinations of the parameters which are invariant under the scaling transformation. This result does not fully agree with that of \cite{Appleby:2012ba}, where the infinite degeneracy does not seem to manifest itself, despite, apparently, allowing all of the parameters to vary. The latter result is not compatible with the existence of the scaling degeneracy of the Galileon equations.

We tried various combinations of data to assess the ability of the Galileon model to fit different observations. In particular, the Galileon model can fit the WMAP9 data better than $\Lambda$CDM with a difference in $\chi^2$ between the best-fitting models of $\Delta\chi^2 \approx -1.8$ (negative values indicate the Galileon model is favoured). However, when the CMB data is combined with measurements from SNIa and BAO, the Galileon model becomes less favored with $\Delta\chi^2 \approx 8.6$. 

We found that the CMB data constrains the Galileon subspace of parameters extremely tightly through the ISW effect; this prediction depends sensitively on the exact values of the parameters. The best-fitting Galileon models are those for which the CMB power spectrum, in the ISW region, decays at low values of $l$. This result is in contrast to $\Lambda$CDM, where the ISW power increases at low $l$ (c.f.~Fig.~\ref{cmb-bf}). We saw that the $c_n$ parameters are mildly correlated with one another (c.f.~Fig.~\ref{galileon-constraints}) and that tiny deviations from the best-fitting regions can quickly lead to a dramatic increase in the ISW power (c.f.~Fig.~\ref{figure-tensions}), rendering the models incompatible with the data. Moreover, the ISW effect is also responsible for the very sharp cut-off in the posterior distribution of the initial energy density of the Galileon field above $\rho_{\varphi, i} / \rho_{m,i} \gtrsim 10^{-6}$ (c.f.~Fig.~\ref{panel-1D}). This dependency of the ISW effect on the Galileon parameters is so strong that adding the low redshift data of SNIa and BAO barely changes the allowed regions of the parameter space (c.f.~Fig.~\ref{galileon-constraints}).

The constraints on the cosmological parameters in the Galileon model, on the other hand, show a much stronger dependence on the combinations of data used. Going from the constraints using WMAP9 data alone to the constraints using the combined WMAP9+SNLS+BAO dataset, we find that the best-fitting values of the total matter density, $\Omega_{m0} h^2$, and amplitude of the primordial scalar fluctuations, $A_s$, shift towards larger values; while the expansion rate $h$, the scalar spectral index of the primordial power spectrum of scalar perturbations, $n_s$, and the optical depth to reioinization, $\tau$, become smaller. The fact that the different datasets prefer different regions of the cosmological parameter space in the Galileon model shows the existence of a tension between the cosmic background evolution and the CMB data, which, ultimately, leads to a poorer fit to the combined datasets.

We compared the constraints on the cosmological parameters in the Galileon model and in $\Lambda$CDM. These two models prefer, approximately, the same best-fitting parameters when using the WMAP9 data alone, with the exception being the expansion rate $h$, which is higher in the Galileon case. However, in the case of the combined WMAP9+SNLS+BAO dataset, the best-fitting parameters of the two models are considerably different, with some of the marginalized two-dimensional posterior distributions being discrepant by more than $2\sigma$ (see Fig.~\ref{cosmological-constraints}). In particular, the Galileon model predicts larger values for $\Omega_{m0}h^2$, $h$ and $A_s$, but lower values for $n_s$ and $\tau$, than $\Lambda$CDM. This shows the importance of varying all the cosmological parameters when comparing to observations, since in this way a broader range of degeneracies in the parameters can be explored, leading therefore to more robust and informative conclusions.

To try to anticipate how the Galileon parameter space can be further constrained by clustering data, we looked at the linear matter power spectra of the best-fitting Galileon models to the combined WMAP9+SNLS+BAO dataset. The situation here is more complicated given the uncertainties about the impact of the highly nonlinear Vainstein screening on the scales for which growth and clustering data are available. In particular, the major issue is whether or not linear theory is still valid on the relevant scales. Nevertheless, provided that linear theory is indeed a good approximation, we have found that the best-fitting models using the combined WMAP9+SNLS+BAO dataset predict too much clustering of matter on small scales, if $b > 1$ (where $b$ is the bias factor). We did manage to find points in the parameter space which predict a linear matter power spectrum closer to $\Lambda$CDM, but the ISW effect for these models is, however, too strong to be compatible with WMAP. Moreover, the best-fitting Galileon models prefer values of $\sigma_8$ which are substantially larger, $\sigma_8 \approx 1$, than in $\Lambda$CDM, where $\sigma_8 \approx 0.8$ (c.f. Table \ref{table-mean}). This suggests that there is a tension between the CMB and matter clustering for this model, which might be difficult to avoid so that adding clustering measurements may significantly tighten the constraints. But again, because of the concerns discussed above, in order to have a cleaner and robust constraint, we prefer to leave the inclusion of clustering and growth data until we have a better understanding of the nonlinear effects in the Galileon model.

We have also looked at the power spectrum of the lensing potential for the best-fitting models constrained by the WMAP9+SNLS+BAO dataset and found them to have, systematically, more power than the $\Lambda$CDM prediction on all angular scales. The currently available weak lensing data is still not good enough to derive meaningful constraints and, therefore, one cannot yet tell if another tension with data will arise from here. Nevertheless, the weak lensing signal seems  to be able to provide a complementary way of distinguishing between the Galileon and the $\Lambda$CDM model. Moreover, the angular scales relevant for the lensing potential power spectrum are large enough for one to assume that the Vainshtein screening effects should be negligible. As a result, it would be interesting to forecast the necessary accuracy of future lensing observations to distinguish between the Galileon model and $\Lambda$CDM.

In summary, we find that the currently available CMB and background data have the power to place very tight constraints on the Galileon parameter space. The model is slightly less favored than $\Lambda$CDM but one cannot yet safely rule it out using data from the CMB and low redshift geometrical tests. On the other hand, the best-fitting Galileon models make very distinctive predictions for the values of $\Omega_{m0}, h, n_s, A_s, \tau$ and $\sigma_8$, which opens for the possibility of the Galileon model to be distinguished from $\Lambda$CDM by higher-significance cosmological data, or by currently available measurements of the galaxy clustering, provided linear theory is applicable. 

\begin{acknowledgments}

We are grateful to Antony Lewis and Lydia Heck for numerical support. We also thank Jonathan Pearson, Gong-Bo Zhao and Miguel Zumalac{\'a}rregui for useful comments and discussions. AB is supported by FCT-Portugal through grant SFRH/BD/75791/2011. BL is supported by the Royal Astronomical Society and Durham University. This work has been partially supported by the European Union FP7  ITN INVISIBLES (Marie Curie Actions, PITN- GA-2011- 289442) and STFC. The Monte Carlo Markov Chains for this paper were run on the ICC Cosmology Machine, which is part of the DiRAC Facility jointly funded by STFC, the Large Facilities Capital Fund of BIS, and Durham University.

\end{acknowledgments}

\bibliography{constraints-galileon.bib}

\end{document}